\documentclass[aps,pre,,twocolumn,preprintnumbers,amsmath,showkeys,showpacs]{revtex4}

\usepackage{lfmath} 
\usepackage{times}
\usepackage{amsmath}
\usepackage{amsfonts}
\usepackage{pstricks,pst-text}

\begin{document}

\title{Locality and stability of the cascades of two-dimensional turbulence}
\author{Eleftherios Gkioulekas}
\email{gkioulekase@utpa.edu}
\affiliation{Department of Mathematics, University of Texas-Pan American , Edinburg, TX, United States}

\begin{abstract}
We investigate and clarify the notion of locality as it pertains to the cascades of two-dimensional turbulence. The mathematical framework underlying our analysis is the infinite system of balance equations that govern the generalized unfused structure functions, first introduced by L'vov and Procaccia. As a point of departure we use a revised version of the system of hypotheses that was proposed by Frisch for three-dimensional turbulence. We show that both the enstrophy cascade and the inverse energy cascade are local in the sense of non-perturbative statistical locality. We also investigate the stability conditions for both cascades. We have shown that statistical stability with respect to forcing applies unconditionally for the inverse energy cascade. For the enstrophy cascade, statistical stability requires large-scale dissipation and a vanishing downscale energy dissipation.  A careful discussion of the subtle notion of locality is given at the end of the paper.
\end{abstract}
\preprint{To be submitted to \emph{Phys. Rev. E}}

\maketitle

\section{Introduction}

The physical notion of locality goes back to the Kolmogorov-Batchelor idea \cite{article:Kolmogorov:1941,article:Kolmogorov:1941:1,article:Batchelor:1947} of an eddy cascade in three-dimensional turbulence where most of the energy is passed on from large eddies to  smaller eddies by cascading through the intermediate scales. The dimensional analysis argument behind the theory of two-dimensional turbulence proposed by Kraichnan \cite{article:Kraichnan:1967:1} , Leith \cite{article:Leith:1968} and Batchelor \cite{article:Batchelor:1969} (KLB) is based in part on the conjecture that a similar physical principle governs the upscale transfer of energy and the downscale transfer of enstrophy. In spite of the importance of the concept of locality to the foundations of the theory of hydrodynamic turbulence, there is no consensus on how to handle the concept rigorously.  The need for a more rigorous understanding of locality becomes more pressing in light of some paradoxical aspects of the theory of two-dimensional turbulence which will be briefly reviewed below. Because quasi-geostrophic models of geophysical flows \cite{article:Charney:1971,article:Salmon:1978,article:Salmon:1980,book:Salmon:1998,article:Orlando:2003:1} relevant both to meteorology and oceonography, and two-dimensional models of magnetically confined plasma turbulence \cite{thesis:Bowman:1992,article:Krommes:1997,article:Krommes:2002} have a similar mathematical structure with two-dimensional turbulence, we cannot simply disregard the paradoxes of two-dimensional turbulence as irrelevant on the grounds that it is a fictitious fluid.

For example, recent numerical simulations \cite{article:Alvelius:2000,article:Kaneda:2001,article:Falkovich:2002,article:Xiao:2003} have validated the KLB prediction $k^{-3}$ for the energy spectrum of the downscale enstrophy cascade. It remains unclear, however, whether the enstrophy cascade is a local cascade or nonlocal cascade. One side of the argument is that it cannot be a local cascade because the slope of the energy spectrum is too steep. On the other hand, if it is not a local cascade, then one has to explain why  the prediction of dimensional analysis agrees with numerical simulations. Furthermore, it is worth remembering that prior to the groundbreaking paper by Lindborg and Alvelius \cite{article:Alvelius:2000}, every attempt to simulate an enstrophy cascade failed. It is now understood that the presence of a dissipation sink at large scales is necessary for a successful simulation of the enstrophy cascade \cite{article:Shepherd:2002,article:Bowman:2003,article:Bowman:2004}. Nonetheless,  we do not have a good grasp on \emph{why} the presence of such a dissipation sink is \emph{sufficient}. A recent theory by Falkovich and Lebedev \cite{article:Lebedev:1994,article:Lebedev:1994:1} predicts the scaling of the logarithmic corrections to the energy spectrum as well as the higher order structure functions of the vorticity for the enstrophy cascade.  However, locality, and the existence of the enstrophy cascade itself are assumptions that are being entered into the theory.  The relevant  question is to understand theoretically the conditions needed for the existence of the enstrophy cascade.

Ironically, the inverse energy cascade presents with an even more confusing situation. From a theoretical standpoint one would not expect the inverse energy cascade to be anything but local. From the standpoint of numerical simulations, there are many positive reports of the predicted $k^{-5/3}$ energy spectrum  \cite{article:Aref:1981,article:Sulem:1984,article:McWilliams:1985,article:Tabeling:1997,article:Tabeling:1998,article:Vergassola:2000}. The most convincing simulation of the inverse energy cascade has been reported in the paper by Boffetta \emph{et al.}\cite{article:Vergassola:2000}, where in addition to the $k^{-5/3}$ prediction, the $3/2$ law has also been confirmed. On the other hand, the locality of the inverse energy cascade has been challenged on the grounds of numerical simulations giving conflicting results \cite{article:Borue:1994,article:Gurarie:2001,article:Gurarie:2001:1,article:Danilov:2003}. The current understanding is that under certain conditions there are coherent structures that spontaneously form while the inverse energy cascade converges to stationarity. Apparently, the inverse energy cascade, as a physical process, continues to take place but it is hidden by the coherent structures which give the dominant contribution to the energy spectrum. Removing the coherent structures artificially by postprocessing simulation data recovers the $k^{-5/3}$ energy spectrum \cite{article:Borue:1994,article:Gurarie:2001:1,article:Fischer:2005}. This aspect of the inverse energy cascade is not well understood. Furthermore, this phenomenon of the spontaneous generation of coherent structures is of considerable interest to  oceonographers.

 In both cases reviewed above the issue at hand is the breakdown of locality. The theoretical challenge is to understand how and why it happens. It should be noted that recent theoretical work \cite{article:Lebedev:1994,article:Lebedev:1994:1,article:Eyink:1995,article:Eyink:1996,article:Yakhot:1999,article:Procaccia:2002,article:Tung:2005,article:Tung:2005:1} that expands on the KLB theory takes locality as well as the existence of the enstrophy cascade and the inverse energy cascade as assumptions. As a result, although various aspects of these cascades have been explained, the more fundamental question of the conditions needed for the existence of the cascades remains elusive.


In the present paper we analyze the locality of the cascades of two-dimensional turbulence by adapting and generalizing the non-perturbative theory of  L'vov  \emph{et al.} \cite{article:Procaccia:1996:1,article:Procaccia:1996:2,article:Procaccia:1996:3,article:Procaccia:1998,article:Procaccia:1998:1,lect:Procaccia:1997}. The mathematical framework is an infinite system of equations that govern the generalized unfused structure functions, the so-called \emph{balance equations}. We also employ a scaling assumption, the \emph{fusion rules}, which we conjecture to be valid in the enstrophy cascade and the inverse energy cascade. The fusion rules govern the scaling of the generalized structure functions when a subgroup of coordinates of velocity differences approach each other. In previous work \cite{article:Tung:2005,article:Tung:2005:1}, we used the balance equations to predict a linear superposition principle between the downscale enstrophy cascade and the hidden downscale energy cascade which exists for finite Reynolds number. In that argument we did not use the fusion rules but we did assume the existence of the cascades. In the present paper we will consider more carefully the implications of the fusion rules on the existence question.


The physical intuition behind our argument is as follows. Let  $F_n$ be the generalized structure function and let $\gz_n$ be its scaling exponent. These structure functions satisfy a system of equations of the form
\begin{equation}
\cO_n F_{n+1} + I_n = \cD_n F_n + Q_n.
\end{equation}
Here, $\cO_n F_{n+1}$ is the nonlinear term that includes the effects of pressure and advection, $I_n$ is a term associated with the sweeping interactions, $Q_n$ is the forcing term, and $\cD_n$ is the dissipation operator. From the fusion rules it can be shown that the integrals in $\cO_n F_{n+1}$ are local under the following conditions: for the downscale cascade UV locality requires $\gz_2 > 0$ and IR locality requires $\gz_{n+1} \leq \gz_2 + \gz_{n-1}$; for the upscale cascade UV locality requires $\gz_n-\gz_{n-2} >0$ and IR locality requires $\gz_{n+1} \geq \gz_2 + \gz_{n-1}$. These conditions can be shown to be satisfied by the \Holder inequalities. It follows that the interactions represented by $\cO_n F_{n+1}$ are local and also self-similar with scaling exponent $\gz_{n+1}-1$. 

The implication of this argument is that the nonlinear interactions accounted for by the term $\cO_n F_{n+1}$ are local both for the enstrophy cascade and for the inverse energy cascade. This notion of locality is called statistical non-perturbative locality \cite{article:Procaccia:1996:3}. However, non-locality, in a different stronger sense,  can arise from the forcing term $Q_n$. Although we may demand that the forcing spectrum be confined to a narrow interval of length scales, it does \emph{not} follow that the forcing term $Q_n$ will force the balance equations only at those length scales. For the case of gaussian forcing, we show that the scaling exponent of $Q_n$ is $q_n=\gz_{n-2}+q_2$ with $q_2 =2$ for the downscale enstrophy cascade and $q_2 <0$ for the inverse energy cascade. It follows that to have true locality we need $q_n-(\gz_{n+1}-1) >0$ in the downscale enstrophy cascade and $q_n-(\gz_{n+1}-1) <0$ in the upscale energy cascade. These conditions are needed for the \emph{statistical stability} of the cascades with respect to forcing perturbations.

It should be noted that nonlocality via the forcing term $Q_n$ is only one of a number of possible scenarios for losing locality. The sweeping term  $I_n$ and the dissipation term $\cD_n F_n$ can also destroy locality under certain conditions. A preliminary discussion of the sweeping term $I_n$ was given in a previous paper \cite{article:Gkioulekas:2007}, and the dissipation term will be discussed in a future publication. Finally, it is also possible to lose locality through violation of the fusion rules. In that case, the term $\cO_n F_{n+1}$ itself would not be local.  In the present paper we will show that the UV locality of the term  $\cO_n F_{n+1}$ is very robust, even under violation of the fusion rules. However the same cannot be said for the IR locality. Our viewpoint then is to consider first the problems that can arise in the favorable case where the fusion rules are valid, before examining the validity of the fusion rules themselves in more depth.

The argument of the present paper supports the conjecture of strong universality \cite{article:Procaccia:2003} for the direct energy cascade of three-dimensional turbulence and the inverse energy cascade of two-dimensional turbulence. However, it definitely rules out strong universality for the downscale enstrophy cascade. Because the argument relies on the hypothesis that the fusion rules hold for the downscale enstrophy cascade and the inverse energy cascade, it is not completely rigorous.  On the other hand, the hypothesis can be investigated by numerical simulation. The $p=2$ fusion rule, which is the essential one with respect to the locality argument, has been proven \cite{lect:Procaccia:1994,article:Procaccia:1995:1,article:Procaccia:1995:2,article:Procaccia:1996} for the direct energy cascade of three-dimensional turbulence, and there is further support by experiments \cite{article:Sreenivasan:1997,article:Procaccia:1998:3,article:Toschi:1998,article:Toschi:1999,article:Reeh:2000,article:Tabar:2000}. For the related problem of the passive scalar \cite{article:Kraichnan:1994}, the fusion rules have been proved for all $p$ \cite{article:Procaccia:2000:1} and have also been confirmed experimentally \cite{article:Procaccia:1996:5,article:Procaccia:1996:6}. The problem of two-dimensional turbulence is similar enough to both problems to make the hypothesis plausible.

From a more philosophical point of view, one can say that the scaling relations implied by the fusion rules are in fact a generalized definition of the physical concept of a ``cascade''.  As has been pointed out previously \cite{article:Procaccia:1996:3}, from a physical standpoint, the fusion rules mean that the large scales are correlated with the small scales in a very particular way where the  self-similarity characteristics of the flow  at the small scales ``forget'' the ongoing physical processes at the large scales (and vice versa for the inverse cascade) which leads to universal scaling. The present argument then establishes the consistency between locality and the scale correlations needed for universality. The conditions needed for this consistency are necessary conditions for the existence of the cascades themselves.

The paper is organized as follows. Section 2 reviews the generalized balance equations of the generalized unfused correlation tensors, the emphasis being on distinguishing the sweeping interactions from the local interactions. Section 3 introduces and motivates our revisions of the Frisch framework of hypotheses as the first step towards a theory of two-dimensional turbulence. The main idea is replacing the anomalous sink hypothesis with a universality hypothesis, which implies the fusion rules hypothesis. In section 4, we extend and generalize the locality proof of L'vov and Procaccia \cite{article:Procaccia:1996:3} to the cascades of  two-dimensional turbulence. In section 5, we then turn to the question of cascade stability, with respect to random gaussian forcing. We find that the inverse energy cascade is stable, but that the enstrophy cascade is only borderline stable, with stability improving as the downscale energy flux is taken to zero. Section  6 discusses various subtleties that arise from our investigation regarding the concept of locality. Some technical matters are relegated to the appendices.

\section{The generalized balance equations}

We now begin by reviewing the theory of the generalized balance equations. These equations were first derived by  L'vov and  Procaccia \cite{article:Procaccia:1996:3} and they are the foundation of previous work  \cite{article:Tung:2005,article:Tung:2005:1,article:Gkioulekas:2007} as well as this paper. The two features of the balance equations that we would like to stress in this paper are the separation of the interaction term into local interactions and sweeping interactions, and the fact that the forcing term can be written in closed form for the case of Gaussian forcing.  We also derive the balance equations that govern the generalized structure functions of the vorticity.

\subsection{Preliminaries}

The governing equations  of two-dimensional turbulence are:
\begin{align}
 \pderiv{u_{\ga}}{t} &+ u_{\gb}\partial_{\gb}u_{\ga} = -\partial_{\ga}p + \cD u_\ga + f_{\ga},\\
\partial_{\ga}u_{\ga} &= 0,
\end{align}
where $f_\ga$ is the forcing term, and $\cD$ is the dissipation operator  given by
\begin{equation}
\cD  \equiv (-1)^{\gk+1} \nu_{\gk}\del^{2\gk}   + (-1)^{m+1}\gb \del^{-2m}.
\end{equation}
Here the integers $\gk$ and $m$ describe the order of the dissipation mechanisms, and  the numerical coefficients $\nu_{\gk}$ and $\gb$ are the corresponding viscosities.  $\cD$ is the overall dissipation operator. The case $\gk = 1$ corresponds to  standard molecular viscosity.  The term $f_\ga$ represents stochastic forcing that injects energy into the system at a range of length scales in the neighborhood of the integral length scale $\ell_0$. The term $\gb \del^{-2m} u_{\ga}$ describes a dissipation mechanism that operates on large-scale motions. The operator $\del^{-2m}$ represents applying the inverse Laplacian $\del^{-2}$ repeatedly $m$ times.  In Fourier space this operator is diagonalized, and its definition may therefore be extended to fractional values for $m$. The same holds for the parameter $\gk$. 

To eliminate pressure we multiply both sides of the Navier-Stokes equation with the operator $\cP_{\ga\gb} \equiv \gd_{\ga\gb} - \pda\pdb\ilapl$ and we employ $\cP_{\ga\gb} u_{\gb} = u_{\gb}$ and $\cP_{\ga\gb} \pdb = 0$ to obtain 
\begin{equation}
\pderiv{u_{\ga}}{t} + \cP_{\ga\gb}\pdc (u_{\gb} u_{\gc}) = \cD u_{\ga} + \cP_{\ga\gb} f_{\gb}.
\end{equation}
 The operator $\cP_{\ga\gb}$ can be expressed in terms of a kernel  $ P_{\ga\gb} (\bfx)$ as 
\begin{align}
\cP_{\ga\gb} v_{\gb} (\bfx) &= \int d\bfy P_{\ga\gb} (\bfx-\bfy)  v_{\gb} (\bfy) \\
 &= \int d\bfy P_{\ga\gb} (\bfy)  v_{\gb} (\bfx-\bfy).
\end{align}
 For two-dimensional turbulence  $ P_{\ga\gb} (\bfx)$ is given by 
 \begin{equation}
 P_{\ga\gb} (\bfx)= \gd_{\ga\gb} \gd (\bfx) - \frac{1}{2\pi}\left[ \frac{\gd_{\ga\gb}}{r^2} - 2 \frac{x_{\ga}x_{\gb}}{r^4} \right].
\end{equation}

The scalar vorticity  $\gz$ is given by $\gz = \gee_{\ga\gb} \pd_\ga u_\gb$ with $\gee_{\ga\gb}$ the Levi-Civita tensor in two dimensions. From the incompressibility condition $\pd_\ga u_\ga = 0$ it follows that there is a function $\gy$, called the streamfunction, such that $u_\ga = \gee_{\ga\gb}\pd_\gb \gy$. Using the identity $\gee _{\ga\gb}\gee _{\gb\gc} = \gd_{\ga\gc}$ one then shows that $\gz = \gee_{\ga\gb}\gee_{\gb\gc}\pd_\ga\pd_\gc \gy = \del^2 \gy$ from which we get $\gy = \del^{-2} \gz$ and $u_\ga = \gee_{\ga\gb} \pd_\gb \del^{-2} \gz$.

The vorticity equation is obtained by differentiating $\gz$ with respect to time and employing the Navier-Stokes equations:
\begin{equation}
\pderiv{\gz}{t} + J(\gy, \gz) = \cD \gz + g,
\end{equation}
where $J(\gy, \gz)$ is the Jacobian defined as
\begin{equation}
J(A, B) = \gee_{\ga\gb} (\pd_\gb A)(\pd_\ga B),
\end{equation}
and $g = \gee_{\ga\gb} \pd_\ga f_\gb$ is the forcing term. The nonlinear term $J \equiv J (\gy, \gz)$ has been obtained by employing the following argument
\begin{align}
J &= \gee_{\ga\gb}\pd_\ga \cP_{\gb\gc}\pd_{\gd} (u_\gc u_\gd ) = \gee_{\ga\gb}\pd_\ga [u_\gc \pd_\gc u_\gb] \\
&= u_\gc \pd_\gc \gz + (\gee_{\ga\gb}\pd_\ga u_\gc) (\pd_\gc u_\gb) \\
&= u_\gc \pd_\gc \gz = J(\gy,\gz).
\end{align}
The term $(\gee_{\ga\gb}\pd_\ga u_\gc) (\pd_\gc u_\gb)$ represents vortex stretching, but in two dimensions it can be shown that 
\begin{equation}
(\gee_{\ga\gb}\pd_\ga u_\gc) (\pd_\gc u_\gb) =0,
\label{eq:notilting}
\end{equation}
 by direct substitution of the vector components.

\subsection{The balance equations}

To write equations  concisely, we introduce the following notation to represent aggregates of position vectors 
\begin{align}
\bfX &= (\bfx , \bfxp), \\ 
\{\bfX \}_n &= \{\bfX_{1} , \bfX_{2} , \ldots , \bfX_{n}\},\\ 
\{\bfX\}_n^k &= \{\bfX_{1} , \ldots , \bfX_{k-1} , \bfX_{k+1} , \ldots , \bfX_{n}\}.
\end{align}
 We use the notation $\{\bfX\}_n  + \gD\bfx$ as a shorthand to represent shifting all the constituent vectors of $\{\bfX\}_n$ by the same displacement $\gD\bfx$. Similarly, $\gl\{\bfX\}_n$ represents taking the scalar product of $\gl$ with every vector in $\{\bfX\}_n$. Finally, the notation $\|\{\bfX\}_n\| \sim R$ means that all point to point distances in the geometry of velocity differences $\{\bfX\}_n$ have the same order of magnitude $R$.  And, the notation $\|\{\bfX\}_n\| \ll \|\{\bfY\}_n\|$ means that all the point to point distances in  $\{\bfY\}_n$ are much larger than all the point to point distances in $\{\bfX\}_n$. 

Let  $w_{\alpha}(\bfx ,\bfxp ,t)$  be the Eulerian velocity differences  
\begin{equation}
w_{\alpha}(\bfx ,\bfxp ,t) = u_{\alpha}(\bfx, t) - u_{\alpha}(\bfxp ,t).
\end{equation}
The eulerian one-time fully unfused correlation tensors are formed by multiplying $n$ velocity differences $w_{\alpha}(\bfx ,\bfxp ,t)$ evaluated at $2n$ distinct points
\begin{equation}
F_n(\{\bfX \}_n, t) = \left\langle \left [  \prod_{k=1}^n w_{\ga_k} (\bfX_{k} , t) \right] \right\rangle.   
\end{equation} 
When all velocity differences share one point in common, that is $\bfxp_k  =\bfx_0$, we say that the correlation $F_n$ is \emph{partially fused}.

The generalized balance equations can be derived by differentiating $F_n$ with respect to $t$ and substituting the Navier-Stokes equations (see appendix \ref{app:balanceproof} for details). This yields the equations
\begin{equation}
\pderiv{F_n}{t} + \cO_n F_{n+1} + I_n = \cD_n F_n + Q_n.
\end{equation}
Here $\cD_n$ is the differential operator representing dissipation, given by
\begin{equation}
\cD_n = \sum_{k=1}^n [\nu (\del_{\bfx_k}^{2\gk} + \del_{\bfxp_k}^{2\gk}) + \gb (\del_{\bfx_k}^{-2m} + \del_{\bfxp_k}^{-2m})],
\end{equation}
and $\cO_n$ is the linear integrodifferential operator such that
\begin{widetext}
\begin{align}
(\cO_n F_{n+1})  (\{\bfx, \bfxp\}_n , t) &= \int O_n (\{\bfX\}_n, \{\bfY\}_{n+1}) F_{n+1} (\{\bfY\}_{n+1}, t) \;  d\{\bfY\}_{n+1} \\
&= \sum_{k=1}^n D_{kn} (\{\bfx, \bfxp\}_n , t) = D_n (\{\bfx, \bfxp\}_n , t),
\end{align}
where $D_{kn}$ is given by
\begin{equation}
D_{kn}^{\ga_1\ga_2\cdots\ga_n} (\{\bfx, \bfxp\}_n , t) = \frac{1}{2n}\sum_{l=1}^n \int d\bfy P_{\ga_k \gb}(\bfy) D_{knl}^{\ga_1\ga_2\cdots\ga_{k-1}\gb\cdots\ga_n} (\{\bfx, \bfxp\}_n , \bfy, t),
\label{eq:locopA}
\end{equation}
with $D_{knl} = D_{knl1} +D_{knl2} + D_{knl3} +D_{knl4}$, and
\begin{align}
D_{knl1}^{\ga_1\cdots\ga_{k-1}\gb\ga_{k+1}\cdots\ga_n} (\{\bfx, \bfxp\}_n , \bfy, t) &= \pd_{\ga_{n+1},\bfx_k} F_{n+1}^{\ga_1\cdots\ga_{k-1}\gb\ga_{k+1}\cdots\ga_{n+1}} (\{\bfX_m\}_{m=1}^{k-1}, \bfx_k - \bfy, \bfxp_k-\bfy, \{\bfX_m\}_{m=k+1}^{n}, \bfx_k - \bfy, \bfx_l), \label{eq:locopB}\\
D_{knl2}^{\ga_1\cdots\ga_{k-1}\gb\ga_{k+1}\cdots\ga_n} (\{\bfx, \bfxp\}_n , \bfy, t) &= \pd_{\ga_{n+1},\bfx_k} F_{n+1}^{\ga_1\cdots\ga_{k-1}\gb\ga_{k+1}\cdots\ga_{n+1}} (\{\bfX_m\}_{m=1}^{k-1}, \bfx_k - \bfy, \bfxp_k-\bfy, \{\bfX_m\}_{m=k+1}^{n}, \bfx_k - \bfy, \bfxp_l),\label{eq:locopC} \\
D_{knl3}^{\ga_1\cdots\ga_{k-1}\gb\ga_{k+1}\cdots\ga_n} (\{\bfx, \bfxp\}_n , \bfy, t) &= \pd_{\ga_{n+1},\bfxp_k} F_{n+1}^{\ga_1\cdots\ga_{k-1}\gb\ga_{k+1}\cdots\ga_{n+1}} (\{\bfX_m\}_{m=1}^{k-1}, \bfx_k - \bfy, \bfxp_k-\bfy, \{\bfX_m\}_{m=k+1}^{n}, \bfxp_k - \bfy, \bfx_l), \label{eq:locopD}\\
D_{knl4}^{\ga_1\cdots\ga_{k-1}\gb\ga_{k+1}\cdots\ga_n} (\{\bfx, \bfxp\}_n , \bfy, t) &= \pd_{\ga_{n+1},\bfxp_k} F_{n+1}^{\ga_1\cdots\ga_{k-1}\gb\ga_{k+1}\cdots\ga_{n+1}} (\{\bfX_m\}_{m=1}^{k-1}, \bfx_k - \bfy, \bfxp_k-\bfy, \{\bfX_m\}_{m=k+1}^{n},\bfxp_k - \bfy, \bfxp_l). \label{eq:locopE}
\end{align}

The term $I_n$ represents the sweeping interactions, and it is given by
\begin{equation}
I_{n}^{\ga_1\ga_2\cdots\ga_n} (\{\bfx, \bfxp\}_n , t) = \sum_{k=1}^n (\pd_{\gc,\bfx_k} + \pd_{\gc,\bfxp_k}) \avg{\cU_{\gc}(\{\bfx_k,\bfxp_k\}_n,t)\left[ \prod_{l=1}^n w_{\ga_l}(\bfx_l, \bfxp_l,t) \right]}.
\end{equation}
\end{widetext}
where $\cU_{\gc}(\{\bfx_k,\bfxp_k\}_n,t)$ is the generalized mean velocity:
  \begin{equation}
  \cU_{\ga} (\{\bfz, \bfzp\}_n , t) = \frac{1}{2n} \sum_{k=1}^n (u_{\ga} (\bfz_k, t) + u_{\ga} (\bfzp_k, t)),
  \end{equation}
The term $Q_n$ represents the forcing term $f_\ga$  and it reads
\begin{equation}
Q_n(\{\bfX\}_n,t) = \sum_{k=1}^n Q_{kn}(\{\bfX\}_n^k,\bfX_k, t),
\label{eq:forceA}
\end{equation}
where $Q_{kn}$ reads
\begin{align}
Q_{kn}^{\ga_1\ga_2\cdots\ga_{n-1}\gb}&(\{\bfX\}_{n-1},\bfY, t) \\
&= \avg{\left[ \prod_{k=1}^{n-1} w_{\ga_k} (\bfX_k, t) \right] \gf_{\gb}(\bfY, t)},
\label{eq:forceB}
\end{align}
with $\gf_{\ga}(\bfX, t) = f_{\ga}(\bfx, t) - f_{\ga}(\bfxp, t)$.

\subsection{Balance equations for the vorticity}
\label{sec:vortbaleqs}

A similar set of equations can be derived for the generalized structure functions of the vorticity.  Let $q (\bfx, \bfxp, t)$ be the vorticity difference defined as 
\begin{align}
q (\bfx, \bfxp, t) &= \gz (\bfx, t) - \gz (\bfxp, t) \\
&= \gee_{\ga\gb} (\pd_{\ga,\bfx} + \pd_{\ga,\bfxp}) w_{\gb} (\bfx, \bfxp, t),
\end{align}
and let $V_n (\{\bfX\}_n, t)$ be the generalized structure function of the vorticity defined as 
\begin{equation}
V_n (\{\bfX\}_n, t) = \avg{\left[\prod_{k=1}^n q (\bfX_k, t)\right]}.
\end{equation}
 It is easy to see that the vorticity generalized structure functions are related to the velocity generalized structure functions by 
\begin{align}
V_n &(\{\bfX\}_n, t)\\
& = \prod_{k=1}^n [\gee_{\ga_k \gb_k} (\pd_{\ga_k,\bfx_k} + \pd_{\ga_k,\bfxp_k})] F^{\gb_1\cdots\gb_n}_n (\{\bfX\}_n, t).
\end{align}
Let  $\cT_n$ be an abbreviation for the differential operator that transforms $F_n$ to  $V_n$ such that $V_n = \cT_n F_n$.

The balance equations for $V_n$ and be derived easily by applying the operator $\cT_n$ on the balance equations for $F_n$. The result is
\begin{equation}
\pderiv{V_n}{t} + \cT_n \cO_n \cT_{n+1}^{-1} V_{n+1} + \cI_n = \cD_n V_n + \cQ_n.
\end{equation}
Here $\cQ_n$ is the forcing term and $\cI_n$ is the sweeping term. The forcing term reads.
\begin{align}
\cQ_n &(\{\bfX\}_n,t) = \sum_{k=1}^n \cQ_{kn}(\{\bfX\}_n^k,\bfX_k, t),\\
\cQ_{kn}&(\{\bfX\}_n^k,\bfY, t) = \avg{\left[ \prod_{k=1}^{n-1} q (\bfX_k, t) \right] g (\bfY, t)}.
\end{align}

To calculate the sweeping term we use \eqref{eq:notilting} to cancel the vortex tilting contributions. With a little bit of algebra we find that 
\begin{widetext}
\begin{align}
\cI_n (\{\bfX\}_n, t) &= \prod_{j=1}^n [\gee_{\ga_j \gb_j} (\pd_{\ga_j,\bfx_j} + \pd_{\ga_j,\bfxp_j})] \sum_{k=1}^n (\pd_{\gc,\bfx_k} + \pd_{\gc,\bfxp_k}) \avg{\cU_{\gc}(\{\bfx_k,\bfxp_k\}_n,t)\left[ \prod_{l=1}^n w_{\gb_l}(\bfx_l, \bfxp_l,t) \right]} \label{eq:VortexSweepingStep} \\
&= \sum_{k=1}^n (\pd_{\gc,\bfx_k} + \pd_{\gc,\bfxp_k}) \avg{\cU_{\gc}(\{\bfx_k,\bfxp_k\}_n,t)\left[ \prod_{l=1}^n  q (\bfx_l, \bfxp_l,t) \right]}.
\end{align}
\end{widetext}

The trick is to apply  the operators $\gee_{\ga_j \gb_j} (\pd_{\ga_j,\bfx_j} + \pd_{\ga_j,\bfxp_j})$ one by one onto the ensemble average in eq. \eqref{eq:VortexSweepingStep}, wherein $n-1$ of the $w_{\gb_l}$ factors are constant for $j\neq l$ with respect to $\bfx_l, \bfxp_l$ , and use the identity $\gee_{\ga\gb}\pd_\ga \pd_\gc [u_\gb u_\gc] = u_\gc \pd_\gc \gz$  on the $w_{\gb_j}$ and $\cU_{\gc}$ factors that are both $\bfx_j, \bfxp_j$ dependent. Each application of these operators  effectively converts each $w_{\gb_l}$ factor into a corresponding $q (\bfx_l, \bfxp_l,t)$ factor. 
The exact mathematical form of the term $\cT_n \cO_n \cT_{n+1}^{-1} V_{n+1}$ is not required.  It is only sufficient to note that once it is shown that the expression $\cO_n \cT_{n+1}^{-1} V_{n+1} = \cO_n F_{n+1}$ is local, then it easily follows that the term $\cT_n \cO_n \cT_{n+1}^{-1} V_{n+1}$  is also local since $\cT_n$ is a linear differential operator.

\section{The theoretical framework}

 Both the K41 theory for three-dimensional turbulence, and the KLB theory for two-dimensional turbulence are based on a dimensional analysis argument.  However, Frisch \cite{article:Frisch:1991,book:Frisch:1995} has suggested that Kolmogorov's second paper \cite{article:Kolmogorov:1941:1} leads to the following more rigorous reformulation of the dimensional analysis argument, based on the following three hypotheses: (H1): At small scales and away from any boundaries, the velocity field is incrementally homogeneous and incrementally isotropic; (H2): Under the same conditions, the velocity field is self-similar at small scales, thereby possessing a unique scaling exponent $h$ ; (H3): the turbulent flow has a non-vanishing mean dissipation rate in the limit of infinite Reynolds number (i.e., an anomalous energy sink).   Then, one uses (H1) and (H3) to derive the 4/5 law which implies that $h=1/3$, and from (H2)  the scaling for all structure functions and the energy spectrum is  deduced. 

In a recent paper, Frisch \cite{article:Frisch:2005} questioned the self-consistency of the assumption of local and incremental homogeneity. The argument essentially is that it is not obvious whether the nonlinearity of the Navier-Stokes equations will preserve incremental homogeneity unless the initial condition is globally homogeneous. In a previous paper \cite{article:Gkioulekas:2007} I have argued that incremental homogeneity will be preserved in the upscale and downscale inertial ranges only if the sweeping interactions, represented by the $I_n$ term of the balance equations, can be neglected in the inertial range. As I have emphasized in that paper, this condition on the $I_n$ term is necessary for the very existence of an inertial range! Here we will simply take it for granted in order to focus our attention on  the other needed conditions.

Within the Frisch framework,  many theoretical approaches to three-dimensional turbulence that try to predict the intermittency corrections to the scaling exponents of the structure functions, can be interpreted as extensions of the Frisch theory where the self-similarity assumption (H2) is weakened   while the other two assumptions (H1) and (H3) are tolerated. It is an easy exercise to reformulate the dimensional analysis argument of the KLB theory in a similar manner.  However, a theory along these lines would already take for granted the locality and universality of the two cascades.   Contrary to the situation in three-dimensional turbulence,  what we must understand \emph{are} the conditions needed to satisfy universality and locality. In previous work \cite{article:Tung:2005,article:Tung:2005:1} we have proposed  that  the questions of locality and universality can be probed more rigorously by adapting the theoretical work of L'vov and  Procaccia \emph{et al.} \cite{article:Procaccia:1996,article:Procaccia:1996:1,article:Procaccia:1996:2,article:Procaccia:1996:3,article:Procaccia:1998,article:Procaccia:1998:1,lect:Procaccia:1997} to two-dimensional turbulence. We will now expand further on this idea on the remainder of the present paper.

\subsection{Revisions to the Frisch framework}

We propose that the Frich framework of hypotheses should be revised as follows:

\emph{ First}, we adopt Frisch's (H1) to our formulation. We have shown previously \cite{article:Gkioulekas:2007} that a stronger homogeneity hypothesis is needed to eliminate the sweeping interactions before deriving the $4/5$ law. Though we may circumvent this problem by postulating that stronger assumption of homogeneity for our framework, we believe that it is desirable to be able to establish the stronger hypothesis from first principles (see section 5 of Ref.\cite{article:Gkioulekas:2007}).

\emph{Second}, to allow for intermittency corrections, it is necessary to relax the  self-similarity hypothesis (H2).  One possibility  is the \emph{space-time self-similarity} assumption, used in the early papers of the quasi-Lagrangian diagrammatic theory \cite{article:Lvov:1987,article:Lvov:1991}.  It was shown later that this assumption is false, because it axiomatically implies Kolmogorov scaling and forbids intermittency corrections \cite{article:Lebedev:1993,article:Procaccia:1998}, thus leading to a self-inconsistent theory. The successful proposal is \emph{space one-time self-similarity}, defined below,  and  we shall adopt it in this paper. 

\emph{Third}, following L'vov and  Procaccia \cite{article:Procaccia:1996:1,article:Procaccia:1996:2,article:Procaccia:1996:3}, we adopt an \emph{hypothesis of universality}. Its purpose is to replace the ad hoc assumption of anomalous sinks.  The universality hypothesis itself claims that statistical symmetries are recovered at length scales away from the forcing range even when the ensemble is constrained by a symmetry-breaking condition at scales closer to the forcing scale.   

Taking the ideas above into consideration, we postulate the following hypotheses for both the enstrophy and energy inertial ranges:
\begin{widetext}


\textbf{Hypothesis 1:} \emph{The velocity field is incrementally stationary, incrementally homogeneous, and incrementally isotropic, defined as }
\begin{gather}
\pderiv{F_n (\{\bfX\}_n, t)}{t} = 0, \forall t\in\bbR, \\
\sum_{k=1}^n (\pd_{\ga_k, \bfx_k} + \pd_{\ga_k, \bfxp_k} ) F_n (\{\bfX\}_n, t) = 0,\\
F_n(\{\bfX\}_n, t) = F_n(\bfr_0 + \cA(\{\bfX\}_n-\bfr_0), t), \;\forall \cA \in SO(2).
\end{gather}
\emph{as long as the evaluations $\{\bfX\}_n$, $\{\bfX\}_n  + \gD\bfr$, $\bfr_0 + \cA(\{\bfX\}_n-\bfr_0)$, lie within an inertial range. }

\textbf{Hypothesis 2:} \emph{The velocity field is self-similar in the sense that for every evaluation $\{\bfX\}_n$ within an inertial range }
\begin{equation}
\exists \gee >0 : F_n(\gl\{\bfX\}_n, t) =\gl^{\gz_n} F_n(\{\bfX\}_n, t), \;\forall \gl\in (1-\gee,1+\gee). 
\end{equation}

For the hypothesis  of universality, we define the conditional correlations
\begin{equation}
\gF_n(\{\bfX\}_n,\{\bfY\}_m,\{\bfw_k\}_{k=1}^{m}, t) = \left\langle  \left. \left [  \prod_{k=1}^n w_{\ga_k} (\bfX_{k} , t) \right]  \right|  \bfw(\bfx_k ,\bfxp_k ,t)=\bfw_k)\right\rangle,
\end{equation}
and use them to formulate the additional hypothesis that in the inertial range, the conditional correlations $\Phi_n$ essentially honor the same symmetries as the unconditional correlations $F_n$, in the asymptotic limit where $\|\{\bfY\}_m\|$ are situated \emph{between} $\|\{\bfX\}_n\|$ and the forcing scale $\ell_0$:

\textbf{Hypothesis 3:} \emph{Let  $\{\bfX\}_n$ and $\{\bfY\}_m$ represent the geometries of velocity differences and let $\cW=\cW(\{\bfY\}_m,\{\bfw_k\}_{k=1}^{m})$.  Then, if in the direct cascade they satisfy $\|\{\bfX\}_n\| \ll \|\{\bfY\}_m\| \ll \ell_0$, or alternatively if in the inverse cascade they satisfy $\|\{\bfX\}_n\| \gg \|\{\bfY\}_m\| \gg \ell_0$,  then the conditional correlations $\gF_n$ preserve incremental stationarity, incremental homogeneity, and incremental isotropy, with respect to $\{\bfX\}_n$, defined as}
\begin{gather}
\pderiv{\gF_n}{t}  = 0, \\
\sum_{k=1}^n (\pd_{\ga_k, \bfx_k} + \pd_{\ga_k, \bfxp_k} ) \gF_n(\{\bfX\}_n, \{\bfY\}_m,\{\bfw_k\}_{k=1}^{m},t) =0\\
\gF_n(\{\bfX\}_n,\{\bfY\}_m,\{\bfw_k\}_{k=1}^{m}, t) = \gF_n(\bfr_0 + \cA(\{\bfX\}_n-\bfr_0),\{\bfY\}_m,\{\bfw_k\}_{k=1}^{m}, t), \;\forall \cA \in SO(2),
\end{gather}
\emph{and also self-similarity, with the same scaling exponents $\gz_n$, defined as}
\begin{equation}
\exists \gee >0 : \gF_n(\gl\{\bfX\}_n,\{\bfY\}_m,\{\bfw_k\}_{k=1}^{m}, t) =\gl^{\gz_n} \gF_n(\{\bfX\}_n,\{\bfY\}_m,\{\bfw_k\}_{k=1}^{m}, t), \;\forall \gl\in (1-\gee,1+\gee). 
\end{equation}
\end{widetext}

Hypothesis 1 is essentially the first hypothesis in the Frisch formulation.  Hypothesis 2 is the \emph{space one-time  self-similarity} principle introduced by L'vov and  Procaccia \cite{article:Procaccia:1996:3} in the context of three-dimensional turbulence. The scaling exponents $\gz_n$ represent the scaling structure of each inertial range. If $0 <\gz_2 < 2$, then the energy spectrum follows a power law given by $E(k)\sim k^{-1-\gz_2}$ \cite{book:Frisch:1995}. If there is a logarithmic correction, then the result also holds  for  $\gz_2 = 2$. Hypothesis 3 states that the statistics of the velocity field at a certain scale still maintain the symmetries stated in hypotheses 1 and 2 even when a symmetry-violating constraint is imposed via a conditional average at scales closer to the forcing scale.  The constituent statements of hypothesis 3 shall be referred to as \emph{universal incremental homogeneity, universal incremental isotropy}, and \emph{universal self-similarity}.  Note that it is essentially  a more careful reformulation of the assumption of ``weak universality'' that was proposed previously by L'vov and  Procaccia \cite{article:Procaccia:1996:1,article:Procaccia:1996:3}. The underlying idea is that the condition $\bfw(\bfx_k ,\bfxp_k ,t)=\bfw_k$ in the definition of the conditional correlations $\Phi_n$  partitions the ensemble of all possible forcing histories consistent with the overall forcing spectrum and the stationarity assumption  into subensembles  defined by the parameters $\{\bfw_k\}_{k=1}^{m}$. Each choice of $\{\bfY\}_m$ represents a distinct partition of the entire ensemble into subensembles. The assumption for the statistical behavior of the velocity field is that it remains invariant accross each subensemble  of forcing histories for all subensemble partitions $\{\bfY\}_m$ (with $\|\{\bfX\}_n\| \ll \|\{\bfY\}_m\| \ll \ell_0$ if it is a downscale cascade or $\|\{\bfX\}_n\| \gg \|\{\bfY\}_m\| \gg \ell_0$ if it is an upscale cascade), and thus dependent only on the overall forcing spectrum.


\subsection{The fusion rules hypothesis}

The immediate consequence of the  universality hypothesis is the fusion rules, whose physical interpretation is that different length scales are correlated (a hint of the cascade process) and that the governing interactions, as we shall show in the next section, are local (a consequence of the fusion rules \emph{and} the structure of the Navier-Stokes equations).  In a forthcoming paper, we will show that the fusion rules also govern the location of the dissipation length scales and  that, in doing so, they provide anomalous energy and enstrophy sinks!

 Consider a geometry of velocity differences $\{\bfX\}_n$ such that all point to point distances have order of magnitude $1$, and define 
\begin{equation}
F_n^{(p)} (r,R) = F_n(r\{\bfX_k\}_{k=1}^{p},R\{\bfX_k\}_{k=p+1}^{n}).
\end{equation}
The function $F_n^{(p)} (r,R)$ reflects the case where $p$ velocity differences have separations with order of magnitude $r$, and $n-p$ velocity differences have separations with order of magnitude $R$. The case of interest is when the evaluation $(r\{\bfX_k\}_{k=1}^{p},R\{\bfX_k\}_{k=p+1}^{n})$ is within the inertial range  $\cJ_n \subseteq (\mathbb{R}^2)^{2n}$ and $r\ll R$.  The fusion rules  give the scaling properties of $F_n^{(p)}$ in terms of the following general form:
\begin{equation}
F_n^{(p)} (\gl_1 r, \gl_2 R) = \gl_1^{\xi_{np}} \gl_2^{\gz_n-\xi_{np}} F_n^{(p)} ( r, R).
\end{equation}
Since $F_n$ is defined as the product of velocity differences we expect the limits $\gl_1\goto 0$ and $\gl_2\goto 0$ to converge.  This implies that $\xi_{np} >0$ and $\gz_n-\xi_{np}>0$ . A concise statement of the fusion rules hypothesis is that for the direct enstrophy cascade $\xi_{np}=\gz_p$ , and for the inverse energy cascade $\xi_{np}=\gz_n-\gz_{n-p}$ for $1<p<n-1$ .  The cases $p=1$ and $p=n-1$ require some additional considerations, and can be deduced, as it turns out, from the $p=2$ fusion rule (see section \ref{sec:crazyfusionrules}). We will also consider the case of ``regular'' violations to the fusion rules where the scaling exponents $\xi_{np}$ satisfy $0<\xi_{np}<\gz_n$, so that the exponents on $\gl_1$ and $\gl_2$ are both positive.

We will now briefly review the argument of L'vov and Procaccia \cite{article:Procaccia:1996:3} that that the fusion rules hypothesis is an immediate consequence of the  universality hypothesis. Let us consider first the case of the direct enstrophy cascade. For the case $2 \leq p \leq n-2$ we will show that for $\nrm{\{\bfX\}_n} \ll \nrm{\{\bfY\}_n}$ the fusion scaling is
\begin{equation}
F_n (\gl \{\bfX\}_p, \mu \{\bfY\}_{n-p}) = \gl^{\gz_p}\mu^{\gz_n-\gz_p} F_n ( \{\bfX\}_p,  \{\bfY\}_{n-p}).
\end{equation}
Let $\cP (\{\bfX\}_n, \{\bfw_k\}_{k=1}^n)$ be the probability for the event  $\bfw(\bfx_k ,\bfxp_k ,t)=\bfw_k$. It follows that
\begin{widetext}
\begin{align}
F_n (\gl \{\bfX\}_p, \mu \{\bfY\}_{n-p}) &= \int \left[\prod_{k=1}^{n-p} w_k\right] \cP  (\mu\{\bfY\}_{n-p}, \{\bfw_k\}_{k=1}^{n-p}) \Phi_p (\gl  \{\bfX\}_p, \mu \{\bfY\}_{n-p}, \{\bfw_k\}_{k=1}^{n-p}) \prod_{k=1}^{n-p} dw_k \\
&= \gl^{\gz_p}  \int \left[\prod_{k=1}^{n-p} w_k\right] \cP  (\mu\{\bfY\}_{n-p}, \{\bfw_k\}_{k=1}^{n-p}) \Phi_p (  \{\bfX\}_p, \mu \{\bfY\}_{n-p}, \{\bfw_k\}_{k=1}^{n-p}) \prod_{k=1}^{n-p} dw_k \\
&= \gl^{\gz_p} F_n (\{\bfX\}_p, \mu \{\bfY\}_{n-p}).
\end{align}
The factor $F_n (\{\bfX\}_p, \mu \{\bfY\}_{n-p})$ is now independent of $\gl$ and has to scale as $\mu^{\gz_n-\gz_p}$.

For  the case of the inverse energy  cascade, again for $2 \leq p \leq n-2$ and under the same limit  $\nrm{\{\bfX\}_n} \ll \nrm{\{\bfY\}_n}$  the fusion scaling is
\begin{equation}
F_n (\gl \{\bfX\}_p, \mu \{\bfY\}_{n-p}) = \gl^{\gz_n-\gz_{n-p}}\mu^{\gz_{n-p}} F_n ( \{\bfX\}_p,  \{\bfY\}_{n-p}).
\end{equation}
We show this with a similar argument as follows:
\begin{align}
F_n (\gl \{\bfX\}_p, \mu \{\bfY\}_{n-p}) &= \int \left[\prod_{k=1}^{p} w_k\right] \cP  (\gl \{\bfX\}_p, \{\bfw_k\}_{k=1}^{p}) \Phi_{n-p} (\mu \{\bfY\}_{n-p}, \gl  \{\bfX\}_p, \{\bfw_k\}_{k=1}^{n-p}) \prod_{k=1}^{p} dw_k \\
&=\mu^{\gz_{n-p}} \int \left[\prod_{k=1}^{p} w_k\right] \cP  (\gl \{\bfX\}_p, \{\bfw_k\}_{k=1}^{p}) \Phi_{n-p} (\{\bfY\}_{n-p}, \gl  \{\bfX\}_p, \{\bfw_k\}_{k=1}^{n-p}) \prod_{k=1}^{p} dw_k \\
&=\mu^{\gz_{n-p}} F_n (\gl \{\bfX\}_p,  \{\bfY\}_{p}).
\end{align}
\end{widetext}
The factor $F_n (\gl \{\bfX\}_p,  \{\bfY\}_{n-p})$  is now independent of $\mu$ and has to scale as $ \gl^{\gz_n-\gz_{n-p}}$.

We would like now to briefly discuss the motivation behind our conjecture that the enstrophy cascade and the inverse energy cascade satisfy the fusion rules.  First, it should be noted that for the locality proof given in section \ref{sec:locality} we only need the fusion rule for the cases $p=2$ and $p=n-2$, from which one then derives the scaling for the cases defined in Fig.~\ref{fig:fusion1} and Fig.~\ref{fig:fusion2}. For the energy cascade of three-dimensional turbulence the $p=2$ fusion rule has been demonstrated  by  Feynman diagram analysis \cite{lect:Procaccia:1994,article:Procaccia:1995:1,article:Procaccia:1995:2,article:Procaccia:1996}.  The proof indicates that the fusion rule essentially follows from the assumption that the scaling exponent $\gz_2$ is universal and does not change in response to perturbations to the forcing statistics.  This assumption rests on less solid ground for the enstrophy cascade, however we can expect it to be true at least in the experimental situations where the cascade actually exists. It is also worth noting that this assumption is weaker than our hypothesis of universality, which in some regard is a stronger assumption than what is really needed. 

There is another consideration that strongly motivates our conjecture: the $p=2$ fusion rule controls the positioning of the dissipation length scale \cite{article:Procaccia:1996:2,article:Procaccia:1996:3,article:Procaccia:1998:3}.  In a forthcoming paper we will show that if this fusion rule is violated, then the dissipation length scale would not be correctly positioned to dissipate the injected energy or enstrophy.  Consequently, it is not easy to reconcile the numerical observation of both cascades with a violation of the fusion rule $p=2$. Furthermore, a situation where the $p=2$ rule is satisfied and the other rules are violated is unlikely. Finally, in two-dimensional turbulence, due to the smaller dimensionality of the problem, we are afforded the opportunity to test of validity of the fusion rules directly with a numerical simulation.

\subsection{Symmetries and the balance equations}

The assumptions that we have put forth  are not self-evident axioms but hypotheses. Thus, the goal of theory is not only to derive conclusions from these assumptions but to also work in the opposite direction and give reasons that justify the assumptions themselves.

The argument that was given by Frisch \cite{article:Frisch:1991,book:Frisch:1995} begins with the observation that the unforced Navier-Stokes equations are invariant with respect to space and time shifts and rotations:
\begin{align}
(t,\bfx, \bfu) &\goto (t, \bfx + \gD \bfx, \bfu), \;\forall \bfx\in\bbR^d, \\
(t,\bfx, \bfu) &\goto (t, A\bfx, A\bfu), \forall A\in SO(d), \\
(t,\bfx, \bfu) &\goto (t+\gD t, \bfx, \bfu).
\end{align}
Furthermore, if we ignore the dissipation terms, then the Navier-Stokes equations are also invariant with respect to the following self-similar transformation
\begin{equation}
(t,\bfx, \bfu) \goto (\gl^{1-h}t, \gl\bfx, \gl^h \bfu), \forall \gl\in\bbR^{+}, h\in\bbR.
\end{equation}
In hydrodynamic turbulence these symmetries are obviously broken by the forcing term, the boundary conditions, and the self-similarity symmetry by the dissipation terms.  Frisch  \cite{article:Frisch:1991,book:Frisch:1995} hypothesized that these symmetries will be statistically reinstated in the inertial range when the flow is governed by a strange attractor. The big question is: how do we prove this? We believe that the generalized balance equations, derived in the previous section, are the proper theoretical framework within which this question can be addressed. 

We begin by accepting the assumption of local stationarity for the reasons given by Frisch \cite{article:Frisch:1991,book:Frisch:1995}.  Then, the balance equations read
\begin{equation}
\cO_n F_{n+1} + I_n = \cD_n F_n + Q_n.
\end{equation}
As was pointed out previously \cite{article:Procaccia:1997,article:Procaccia:1998}, the advantage of using generalized structure functions where every velocity difference is associated with two distinct coordinates that are different from any other velocity difference, is that in the limits $\nu\goto 0$ and $\gb\goto 0$ the dissipation terms can be dropped. This is not possible for the standard structure functions where every velocity difference is associated with the same two coordinates. We show this by using the mean-value theorem to bound $\cD_n F_n$ as follows
\begin{equation}
|\cD_n F_n | \leq \left(\frac{C_1 \nu}{R_{\min}^{2\gk}} + C_2\gb R_{\max}^{2m}\right) | F_n |.
\end{equation}
Here, $C_1$ and $C_2$ are constants independent of $\nu$ and $\gb$, and
\begin{align}
R_{\min} &\equiv \min \{ \bfx_k, \bfxp_k : k\in \bbN, 1\leq k\leq n \}, \\
R_{\max} &\equiv \max \{ \bfx_k, \bfxp_k : k\in \bbN, 1\leq k\leq n\}. 
\end{align}
It is easy to see that because all the differentiations can be performed without invoking the product rule, the viscosities $\nu$ and $\gb$ multiply on a factor that remains finite in the limits $\nu\goto 0$ and $\gb\goto 0$. Thus,  $|\cD_n F_n | \goto 0$ in the inertial range.

To reinstate the statistical symmetries we need a region of length scales where $Q_n$ and $I_n$ can also be ignored.  Then, one has the homogeneous equation $\cO_n F_{n+1}=0$,  which remains invariant both under local homogeneity and local isotropy. In fact, it is also known \cite{article:Procaccia:1998,article:Procaccia:1998:1} that the homogeneous equation is invariant under the following similarity transformation: 
\begin{equation}
\{\bfX\}_n \goto \gl\{\bfX\}_n, \quad F_n \goto \gl^{nh+\cZ (h)} F_n.
\end{equation}
consequently, it is expected to have solutions in the general form   
\begin{equation}
F_n = \int d\mu (h) F_{n,h},
\end{equation}
 where $F_{n,h}$ are the zero-modes of the operator $\cO_n$ which scale as
\begin{equation}
F_{n,h} (\gl \{\bfX\}_n, t) = \gl^{nh+\cZ (h)} F_{n,h} (\{\bfX\}_n, t).
\end{equation}
Note that the same result can also be obtained from the multifractal hypothesis \cite{book:Frisch:1995}. On a large inertial range, the leading contribution to $F_n$ is asymptotically self-similar with the scaling exponent $\gz_n$ given by 
\begin{equation}
\gz_n = \min_h (nh+\cZ (h)).
\end{equation}
For the case of a multifractal stochastic velocity field with  $D (h)$ defined as the fractal dimension of the set of points that support a local \Holder exponent $h$, the relationship between $\cZ (h)$ and $D (h)$ is $\cZ (h) = d - D (h)$ where $d$ is the dimension of the velocity field, and $d=2$ for two-dimensional turbulence.

It has been suggested, for the case of three-dimensional turbulence, that the scaling exponents $\gz_n$ can be calculated from the solvability condition of the homogeneous equation $\cO_n F_{n+1} = 0$  \cite{article:Procaccia:1998,article:Procaccia:1998:1,article:Procaccia:1998:2}. Although, from a practical standpoint,  perturbative methods have been more effective \cite{article:Procaccia:2000,article:Giles:2001}, the solvability condition analysis reveals the underlying principle governing the origin of the scaling exponents $\gz_n$. From a physical standpoint, the condition $\cO_n F_{n+1} = 0$ includes (for $n=2$) \emph{and extends} (for $n>2$) the requirement of a ``constant'' (in the asymptotic sense) energy flux  in the inertial range. The extension makes the condition powerful enough to lock down all the scaling exponents $\gz_n$, as was demonstrated by Belinicher \emph{et al.} \cite{article:Procaccia:1998:2}.  As we have shown in a previous paper \cite{article:Tung:2005}, the problem with extending this argument to two-dimensional turbulence is that the scaling exponents $\gz_n$ of the enstrophy cascade are not non-trivial solutions to the equation $\cO_n F_{n+1} = 0$. This should not surprise us, that we cannot obtain the scaling exponents of the enstrophy cascade from an ``extended'' constant energy flux condition! What must be done instead is to use the equation
\begin{equation}
\cT_n\cO_n F_{n+1} = 0,
\label{eq:homeq}
\end{equation} 
obtained by the generalized balance equations for the vorticity structure functions derived previously in section \ref{sec:vortbaleqs}. This equation represents an ``extended'' constant \emph{enstrophy} flux condition, and it yields two solutions for the scaling exponents, instead of just one: an energy cascade solution that transfers energy but not enstrophy (the non-trivial solution of $\cO_n F_{n+1} = 0$ and it also satisfies $\cT_n\cO_n F_{n+1} = 0$ trivially because it does not transfer enstrophy), and an enstrophy cascade solution that transfers enstrophy but not energy (the non-trivial solution of $\cT_n \cO_n F_{n+1} = 0$ and it also satisfies $\cO_n F_{n+1} = 0$ trivially because it transfers no energy).  It also follows from the mathematical structure of the equation \eqref{eq:homeq} that these two solutions can be superimposed linearly to obtain a composite solution that transfers both energy and enstrophy. The possibility and implications of such a composite solution  has been discussed in previous papers \cite{article:Tung:2005,article:Tung:2005:1,article:Tung:2007,article:Tung:2006}, and will not concern us further in this paper.

These observations show that a constructive point of view is to see  our hypotheses 1,2, and 3  as an efficient \emph{definition} of the concept of an ``inertial range'', in a generalized sense. Obviously, the  hypotheses are valid only on a multidimensional domain of velocity differences geometries $\{\bfX\}_n \in \cJ_n$.  The extent of this domain $\cJ_n$ \emph{is}  the extent of the inertial range itself. A one-dimensional interval of length scales  where the structure functions exhibit power law scaling,  is a  reduction of the domain $\cJ_n$ in which information is lost.  For the case of dual cascade, we have an upscale range and a downscale range, and a different set of scaling exponents $\gz_n$ and region $\cJ_n$ is associated with each range. To determine the extent of the region $\cJ_n$ for the energy and enstrophy ranges we  employ the theory of the generalized balance equations, combined with the fusion rules hypothesis.  More rigorously, the domain $\cJ_n$ is the range of length scales where the terms $Q_n$, $I_n$, and $\cD_n F_n$ in the generalized balance equation are negligible relative to  the terms contributing to $\cO_n F_{n+1}$.  The first step towards determining the extent of the domain $\cJ_n$ is to calculate, from our hypotheses, the scaling exponents of the terms of the balance equations. Then these terms can be compared against each other. We initiate this study in the next two sections of this paper. Note that it is sufficient to study in this manner  only the balance equations for the velocity field. Since the operator $\cT_n$ is a strictly differential operator, it is also local, therefore the scaling exponents of the terms of the vorticity  balance equations are equal to the  the scaling exponents of the terms of the velocity balance equations minus 1. So, the scaling exponents for pairwise ratios of the terns against each other are the same for both balance equations.


\section{Locality of the interaction term}
\label{sec:locality}

\psset{unit=10pt,labelsep=5pt}

\begin{center}
\begin{figure}[t]
\scalebox{1}{\includegraphics{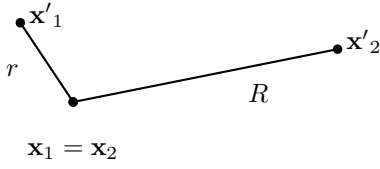}}
\caption{\label{fig:fusion1}\small The $p=1$ fusion rule geometry with a type B fusion. Here we take the limit $r \ll R$ with $r$ and $R$ both in the inertial range. In a type B fusion, the small velicity difference shares an endpoint with one of the large ones, i.e. $\bfx_1 =\bfx_2$.}
\end{figure}
\end{center}

\begin{figure}[t]\begin{center}
\scalebox{1}{\includegraphics{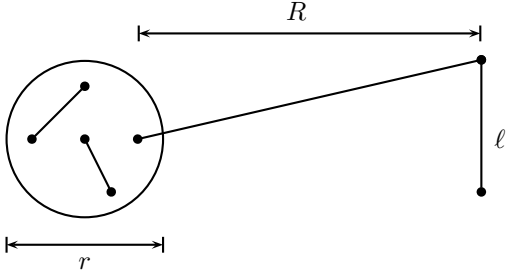}}
\caption{\label{fig:fusion2}\small The $p=n-1$ fusion rule geometry with a type B fusion. This is a composite rule where we take the limits $\ell \ll R$ and $r \ll R$. The velocity difference associated with $\ell$ shares an endpoint with the velocity difference associated with $R$.}
\end{center}\end{figure}

We will now show that the $p=2$ fusion rule and $p=n-2$ fusion rule combined with local homogeneity and incompressibility, implies that the nonlinear interactions in the inertial range are local. From the viewpoint of the generalized balance equations, the nonlinear interactions are accounted for by the integral in the term $\cO_n F_{n+1}$, and the sweeping interactions by the term $I_n$, which we assume, for now, that it is negligible in the inertial range (see Ref. \onlinecite{article:Gkioulekas:2007}  for further discussion). We say that the integral is \emph{local} if it is convergent and furthermore if the dominant contribution to the integrals in $\cO F_{n+1}$ comes from the region in which the separation of the integral variable $\bfy$ from all other points has the same order of magnitude as $\{\bfX\}_n \sim R$. Locality implies that  the contributions $D_{kn}$ to $\cO F_{n+1}$ are also self-similar with scaling exponent $\gd_n$ and satisfy
\begin{equation}
D_{kn}(\gl\{\bfX\}_n, t) =\gl^{\gd_n} D_{kn}(\{\bfX\}_n, t),
\end{equation}
where $\gd_n$ is given by $\gd_n = \gz_{n+1} - 1$. \emph{We propose that the locality of the interaction integral in $D_{kn}$ is the mathematical definition that corresponds most closely to the kind of locality that is required to enable an eddy  cascade  with universal scaling.}  In the sense of our proposed definition, we will show that both the energy and enstrophy cascade of two-dimensional turbulence \emph{are} local. 

The proof given in this section is based on a previous proof by L'vov and Procaccia given in  section IV-C of Ref.  \onlinecite{article:Procaccia:1996:3}.  The same argument is also presented in the appendix of Ref. \onlinecite{article:Procaccia:1997}. We have generalized their proof in two directions: first, we derive the explicit conditions needed for locality even for the case where the fusion rules do not hold; second, we extend the proof to the case of the inverse energy cascade.

\subsection{Preliminaries}
\label{sec:crazyfusionrules}

It can be seen from the equations \eqref{eq:locopA},  \eqref{eq:locopB}, \eqref{eq:locopC},  \eqref{eq:locopD},  \eqref{eq:locopE} that the general form of the terms that contribute to $D_{kn}$ involves an integral of the form
\begin{widetext}
\begin{equation}
 \cI = \int d\bfy\; P_{\ga_k\gb}(\bfy) \partial_{\gc,\bfx_k}\avg{\left[\prod_{l=1,l\neq k}^n w_{\ga_l} (\bfX_l)\right] w_{\gb}(\bfx_k-\bfy,\bfxp_k-\bfy) w_{\gc}(\bfx_k-\bfy, \bfs)},
\label{eq:theintegral}
\end{equation}
\end{widetext}
where $\bfs$ can be any point among $\bfx_1,\ldots,\bfx_n$ or $\bfxp_1,\ldots,\bfxp_n$. The locality proof requires the scaling of $F_n$ in the  limits   $\bfy\goto\bfzero$, $\bfx_k-\bfy\goto \bfx_l \text{ or } \bfxp_l$,  $\bfxp_k-\bfy\goto \bfx_l \text{ or } \bfxp_l$, and $\gr=\nrm{\bfy} \goto \infty$. Consequently, we need the fusion rules for the geometries shown in Fig. \ref{fig:fusion1} (case $p=1$) and Fig. \ref{fig:fusion2} (case $p=n-1$). Both can be derived from the fusion rules for the cases $p=2$ and $p=n-2$.

(a) For the case $p=1$ where we also  assume  a type 1B fusion (i,e.   $\bfx_1 = \bfx_2$, and see  Fig. \ref{fig:fusion1}) the governing fusion rule is 
\begin{align}
F_n &\sim (r/R)^{\gz_2} R^{\gz_n} \text{ (downscale)},\\
F_n &\sim (r/R)^{\gz_n-\gz_{n-2}} R^{\gz_n} \text{ (upscale)}.
\end{align}
To show this,  we note that
\begin{align}
\bfw (\bfx_2, \bfxp_2) &= \bfw (\bfx_2, \bfxp_1) + \bfw (\bfxp_1, \bfxp_2) \\ 
&=  \bfw (\bfx_1, \bfxp_1) + \bfw (\bfxp_1, \bfxp_2).
\end{align}
For the last step, we used $\bfx_1 = \bfx_2$. Let $Y=(\bfxp_1, \bfxp_2)$. Then
\begin{align}
F_n (\{\bfX\}_n) &= F_n (\bfX_1, \bfX_2, \{\bfX\}_{k=3}^n)\\
&= F_n (\bfX_1, \bfX_1, \{\bfX\}_{k=3}^n) + F_n (\bfX_1, \bfY, \{\bfX\}_{k=3}^n).
\end{align}
The third term is the same fusion problem as the first term because $\bfX_1$ and $\bfY$ share the point $\bfxp_1$, and from the universal isotropy hypothesis we can rotate the legs $r$ and $R$ in Fig. \ref{fig:fusion1} with respect to each other so that the three points form an isosceles triangle. Then one problem can be obtained from the other problem by reflection around the triangle's axis of symmetry. Consequently,  both problems scale according to the second term, which is a $p=2$ fusion. In the proof below, we will use the generalized scaling
\begin{equation}
F_n \sim (r/R)^{\xi_{n,2}} R^{\gz_n},
\end{equation}
 which is applicable both upscale and downscale.

(b) For $p=n-1$ with type B fusion, we have $n-2$ velocity differences of order $r$, one  velocity difference  of order $\ell$ with one endpoint attached to a velocity difference  of order $R$, where $\ell \ll R$ and $r \ll R$.   Note that this fusion can be composed as follows. Begin with all velocity differences at order $R$. Then take the following limits: ($\ell_1$) Shrink one velocity difference to order $\ell \ll R$ with one endpoint attached to another velocity difference (this is the previous case); ($\ell_2$) Shrink all other $n-2$ velocity differences down to order $r\ll R$. Thus, we have, for the downscale case,
\begin{align}
F_n &\sim \fracp{\ell}{R}^{\gz_2}\fracp{r}{R}^{\gz_{n-2}} R^{\gz_n}\\
& \sim \ell^{\gz_2} r^{\gz_{n-2}} R^{\gz_n-\gz_{n-2}-\gz_2}.
\end{align}
The first limit ($\ell_1$) gives the first factor $(\ell/R)^{\gz_2}$, and the second limit ($\ell_2$) the second factor $(r/R)^{\gz_{n-2}}$. Similarly, for the upscale case, using the exact same limits $(\ell_1)$ and $(\ell_2)$ , we find
\begin{align}
F_n &\sim \fracp{\ell}{R}^{\gz_n-\gz_{n-2}}\fracp{r}{R}^{\gz_n-\gz_{2}} R^{\gz_n} \\
&\sim \ell^{\gz_n-\gz_{n-2}} r^{\gz_n-\gz_{2}} R^{\gz_2+\gz_{n-2}-\gz_n}.
\end{align}
In the proof below, we will use the generalized scaling
\begin{equation}
F_n \sim \fracp{\ell}{R}^{\xi_{n,2}}\fracp{r}{R}^{\xi_{n,n-2}} R^{\gz_n}.
\end{equation}

\subsection{UV locality}

UV locality requires convergence in the limits  $\bfy\goto\bfzero$, $\bfx_k-\bfy\goto \bfx_l \text{ or } \bfxp_l$, and $\bfxp_k-\bfy\goto \bfx_l \text{ or } \bfxp_l$. The only limit that requires serious consideration is the first where $P_{\ga\gb}(\bfy)$ is singular. For this case we distinguish the following two subcases.

(a) Assume that $\bfx_k \neq \bfs$. The derivative of the ensemble average in \eqref{eq:theintegral}  is analytic in $\bfy\goto\bfzero$, so we Taylor expand it around $\bfy=0$. 
\begin{equation}
\cI = \int d\bfy\; P_{\ga_k\gb} (\bfy) [\cA_{\gb} + \cB_{\gb\gc}\bfy_{\gc} + \cC_{\gb\gc\gd}\bfy_{\gc}\bfy_{\gd} + \cdots ].
\end{equation}
The first term vanishes by incompressibility. The second term vanishes because the integral is odd with respect to $\bfy$, from the local isotropy hypothesis, whereas $P_{\ga_k\gb}(\bfy)$ is even. The third integral is local. Use $d\bfy = \gr \;d\gr \;d\Omega (A)$ with $\gr = \nrm{\bfy}$, $A \in SO (2)$, and $d\Omega (A)$ the measure of two-dimensional spherical integration. The third integral then reads
\begin{align}
\cI_3 &= \int d\gr \int d\Omega (A) \gr P_{\ga_k\gb}(\bfy) \cC_{\gb\gc\gd}\bfy_{\gc}\bfy_{\gd} \\
&\sim \int_{0^{+}} d\gr\; \gr \gr^{-2} \gr^2  \sim \int_{0^{+}} d\gr\; \gr \sim \gr^2,
\end{align}
and it is unconditionally local

\begin{figure}[tb]\begin{center}
\scalebox{1}{\includegraphics{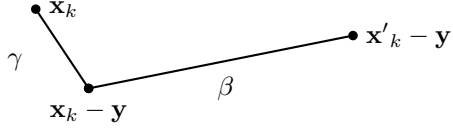}}
\caption{\label{fig:locality-uv}\small UV limit for the case $\bfx_k = \bfs$. We employ the fusion rule shown in Fig.~\ref{fig:fusion1}}
\end{center}\end{figure}

(b) Assume $\bfx_k = \bfs$. Then the integral reads
\begin{widetext}
\begin{equation}
 \cI = \int d\bfy\; P_{\ga_k\gb}(\bfy) \partial_{\gc,\bfx_k}\avg{\left[\prod_{l=1,l\neq k}^n w_{\ga_l} (\bfX_l)\right] w_{\gb}(\bfx_k-\bfy,\bfxp_k-\bfy) w_{\gc}(\bfx_k-\bfy, \bfx_k)},
\end{equation}
\end{widetext}
and in the limit $\bfy\goto\bfzero$ we have the velocity difference geometry shown in Fig.~\ref{fig:locality-uv}. From the $p=1$ fusion rule with type 1B fusion, the ensemble average in the integral scales as $F_{n+1} \sim (\gr/R)^{\xi_{n+1,2}} R^{\gz_{n+1}}$. The integral then scales as
\begin{equation}
\cI \sim \int_{0^{+}} d\gr \;\gr \gr^{-2}\gr^{-1}\gr^{\xi_{n+1,2}} \sim \int_{0^{+}}d\gr\; \gr^{\xi_{n+1,2}-2}.
\end{equation}
Here, the spherical integral contributes the factor $\gr$, the projection operator $P_{\ga_k\gb}(\bfy)$ contributes $\gr^{-2}$, the derivative $\partial_{\gc,\bfx_k}$ contributes $\gr^{-1}$ (because the $\bfx$ dependent factor depends only on the smallest in separation  of the two velocity differences in Fig.~\ref{fig:locality-uv}, which makes that factor dependent only on $\gr$), and the fusion rule contributes $\gr^{\xi_{n+1,2}}$. The resulting integral is marginally local for $\xi_{n+1,2}=\gz_2 =2$ (enstrophy cascade) and non-local for $\xi_{n+1,2}=\gz_2 = 2/3$ (downscale energy cascade in 3D).  However, note that  the type 1B fusion rule for the case $p=1$, which we have used here, is written in more detail as:
\begin{align}
F_{n+1} &\sim \avg{w_{\gb} (\bfx_k-\bfy, \bfx_k)w_{\gc}(\bfx_k-\bfy, \bfx_k)} \Phi_{n-1}\\
&\sim \Phi_2 (\bfx_k-\bfy, \bfx_k,\bfx_k-\bfy, \bfx_k) \Phi_{n-1},
\end{align}
which allows the integral $\cI$ to be rewritten as
\begin{equation}
\cI \sim \Phi_{n-1}  \int d\bfy\; P_{\ga_k\gb}(\bfy) \partial_{\gc,\bfx_k} \Phi_2 (\bfx_k-\bfy, \bfx_k,\bfx_k-\bfy, \bfx_k).
\end{equation}
Here we have used the fact that $\Phi_{n-1}$ is independent of both $(\bfx_k-\bfy, \bfx_k)$ and $(\bfx_k-\bfy,\bfxp_k-\bfy)$, thus independent of $\bfx_k$, and therefore it can be pulled out of the $\partial_{\gc,\bfx_k}$ operator. It is easy to see that the leading term of the $\Phi_2$ factor  vanishes when differentiated by $\partial_{\gc,\bfx_k}$ by universal incremental homogeneity.  Thus, we get a cancellation that kills the leading contribution and  the integral then scales according to the next-order term:
\begin{equation}
\cI \sim  \int_{0^{+}}d\gr \;\gr^{\xi_{n+1,2}-1} \sim\gr^{\xi_{n+1,2}}.
\end{equation}
This integral  is \emph{local} if $\xi_{n+1,2} > 0$ (i.e. for locality we need $\cI \goto 0$ as $\gr \goto 0$). The result holds unconditionally, even under a regular violation of the $p=2$ fusion rule, e.g.  $F_{n+1} \sim (\gr/R)^{\xi_{n+1,2}} R^{\gz_{n+1}}$ as long as $\xi_{n+1,2} > 0$ and some factorization $F_{n+1}\sim\Phi_2\Phi_{n-1}$ is still possible (that would be true for higher-order terms, if the leading term should happen to vanish) . Under the fusion rules hypothesis this condition is $\gz_2 > 0$ for a downscale cascade and $\gz_{n+1}-\gz_{n-1} > 0, \forall n\in\bbN-\{0,1\}$ for an upscale cascade.

Consider finally the cases $\bfx_k-\bfy\goto \bfx_l \text{ or } \bfxp_l$, and $\bfxp_k-\bfy\goto \bfx_l \text{ or } \bfxp_l$. We perform the integral spherically around the value of $\bfy$ where one of these coincidences take place. Let $\gr$ be the distance between the two approaching points. Assume any regular fusion rule of the form  $F_{n+1} \sim (\gr/R)^{\xi_{n+1,2}} R^{\gz_{n+1}}$. Now, the function $P_{\ga_k\gb}(\bfy) $ is no longer singular so we gain a factor of $\gr^2$. Otherwise, the computation is the same as in the previous case, and the integral scales as 
\begin{equation}
\cI \sim  \int_{0^{+}}d\gr\; \gr^{\xi_{n+1,2}+1} \sim \gr^{\xi_{n+1,2}+2},
\end{equation}
which is  local even under a regular violation of the $p=2$ fusion rule.

\subsection{IR locality}

\begin{figure}[tb]\begin{center}
\scalebox{1}{\includegraphics{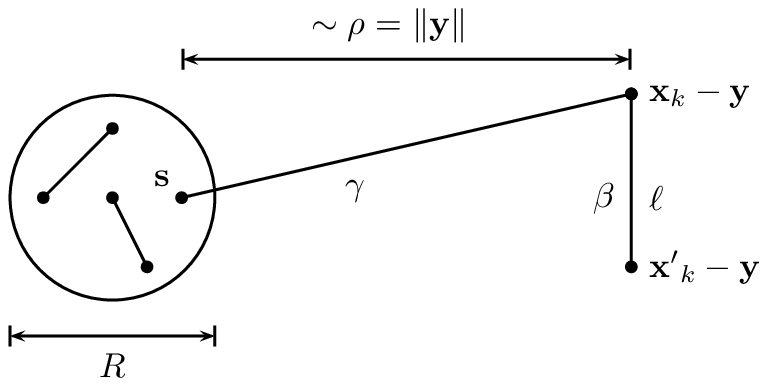}}
\caption{\label{fig:locality-ir}\small IR limit $\nrm{\bfy}\goto\infty$. We employ the fusion rule shown in Fig.~\ref{fig:fusion2}}
\end{center}\end{figure}

Consider the limit $\gr=\nrm{\bfy} \goto \infty$. The corresponding geometry of velocity differences is shown in Fig.~\ref{fig:locality-ir}. For the downscale cascade we use the fusion rule for the case $p=n-1$, defined in Fig. \ref{fig:fusion2}:
\begin{equation}
F_{n+1} \sim \fracp{\ell}{\gr}^{\xi_{n+1,2}}\fracp{R}{\gr}^{\xi_{n+1,n-1}} \gr^{\gz_{n+1}}.
\end{equation}
Expanding around the point at infinity $\gr\goto\infty$, we get the asymptotic expansion
\begin{equation}
F_{n+1} \sim \gr^{\gz_{n+1}-\xi_{n+1,2} - \xi_{n+1,n-1}}(c_0 + c_1 \gr^{-1} + c_2 \gr^{-2}+\cdots).
\end{equation}
The integral then scales as
\begin{equation}
\cI \sim \int^{\infty} d\gr\; \gr\gr^{-2}\gr^{\gz_{n+1}-\xi_{n+1,2} - \xi_{n+1,n-1}}(c_0 + c_1 \gr^{-1}+\cdots).
\end{equation}
 Here, the spherical integral contributes the factor $\gr$, and the projection operator contributes $\gr^{-2}$.  In this limit, the derivative $\partial_{\gc,\bfx_k}$ does not contribute a factor of $\gr^{-1}$, because the only factor that can be $\bfx_k$ dependent is the factor that gives $(\ell)^{\xi_{n+1,2}}$. This factor is dependent on $\ell$ and independent of $\gr$, again because $\ell$ is the smallest distance. On the other hand, the effect of the derivative $\partial_{\gc,\bfx_k}$ is to vanish the $\Phi_2$ factor altogether via an incompressibility cancellation.  To see this, note that the fusion rule corresponding to the geometry of Fig.~\ref{fig:locality-ir} gives
\begin{align}
F_{n+1} &\sim \avg{w_{\gb} (\bfx_k-\bfy, \bfxp_k-\bfy)w_{\gc}(\bfx_k-\bfy, \bfxp_k-\bfy)} \Phi_{n-1}\\
&\sim \Phi_2 (\bfx_k-\bfy, \bfxp_k-\bfy,\bfx_k-\bfy, \bfxp_k-\bfy) \Phi_{n-1},
\end{align}
and from the incompressibility condition we get the tensor structure of $\Phi_2$  which is
\begin{equation}
\Phi_2 \sim \left[ (2+\xi_{n+1,2}) \gd_{\gb\gc} - \xi_{n+1,2} \frac{\ell_\ga \ell_\gb}{\ell^2} \right] \ell^{\xi_{n+1,2}},
\end{equation}
with $\ell = \nrm{\bfx_k - \bfxp_k}$. The integral $\cI$ can be rewritten as
\begin{equation}
\cI \sim \Phi_{n-1}  \int d\bfy\; P_{\ga_k\gb}(\bfy) \partial_{\gc,\bfx_k} \Phi_2 (\bfx_k-\bfy, \bfxp_k-\bfy,\bfx_k-\bfy, \bfxp_k-\bfy).
\end{equation}
Again, $\Phi_{n-1}$ is independent of $\bfx_k$ and can be pulled out of the derivative $\partial_{\gc,\bfx_k}$. However, differentiating with respect to $\bfx_k$ wiggles only one of two points (that is $\bfx_k-\bfy$, but not $\bfxp_k-\bfy$),  which makes it, by chain rule, a derivative with respect to $\ell$, which in turn vanishes due to  the tensor structure of $\Phi_2$ above. As a result, we pick the factor $c_1 \gr^{-1}$ from the next order term, and the integral scales as:
\begin{align}
\cI &\sim \int^{\infty} d\gr\; \gr^{\gz_{n+1}-\xi_{n+1,2} - \xi_{n+1,n-1}-1} c_1 \gr^{-1} \\
&\sim \gr^{\gz_{n+1}-\xi_{n+1,2} - \xi_{n+1,n-1}-1}.
\end{align}
The locality condition for this integral is $\gz_{n+1}-\xi_{n+1,2} - \xi_{n+1,n-1} \leq 0$ and thus $\gz_{n+1} \leq \xi_{n+1,2}  + \xi_{n+1,n-1}$. For a downscale cascade, the  fusion rules hypothesis gives the condition $\gz_{n+1} \leq \gz_2  + \gz_{n-1}$. For an upscale cascade, the  fusion rules hypothesis reads $\xi_{np}=\gz_n-\gz_{n-p}$, therefore the condition now reads $\gz_{n+1} \geq \gz_2  + \gz_{n-1}$. The condition for locality is the same as in the downscale cascade, but the direction of the inequality is reversed.

\subsection{Summary}

Let us now summarize what has been proved.  We have shown that for either a downscale or an upscale cascade the locality conditions are 
\begin{align}
\text{(UV): }&\xi_{n+1,2} > 0, \;\forall n\in \bbN, n>1  \\
\text{(IR): }&\gz_{n+1} \leq \xi_{n+1,2} + \xi_{n+1,n-1}, \;\forall n\in \bbN, n>1.
\end{align}
for UV locality and  IR locality correspondingly.  For a downscale cascade,  the IR locality  condition is satisfied under the fusion rules hypothesis 
\begin{equation}
\xi_{np}=\gz_p,\; \forall p,n\in\bbN, n>1,  2\leq p \leq n-2,
\end{equation}  
due to the \Holder inequality $\gz_{n+1} \leq \gz_2+\gz_{n-1}$ for the scaling exponents $\gz_n$.  For an upscale cascade, the  fusion rules hypothesis gives  
\begin{equation}
\xi_{np} = \gz_n-\gz_{n-p}, \; \forall p,n\in\bbN, n>1,  2\leq p \leq n-2,
\end{equation} 
 and the IR locality condition is reduced to $\gz_n \geq \gz_2+\gz_{n-2}$ which is still satisfied, because the \Holder inequality reverses its direction when the cascade is upscale (see appendix \ref{sec:holder}). The UV locality  condition is also satisfied, but does not require the fusion rules hypothesis. All that is required is that the scaling exponent $\xi_{n,2}$ be positive.  For a downscale cascade this gives the condition $\gz_2 > 0$ and for an upscale cascade,  the condition $\gz_{n+1}-\gz_{n-1} > 0$. The assumption of the regular fusion scaling is sufficient for that, for both upscale and downscale cascades.

Let us now consider the case where the fusion rules are violated according to 
\begin{align}
\xi_{np} &= \gz_p+\gD \xi_{np} \text{ (downscale)},\\
\xi_{np} &= \gz_n-\gz_{n-p}+\gD \xi_{np} \text{ (upscale)}.
\end{align}
As we have argued above, as long as the violation is regular, UV localiy is still maintained.  For IR  locality, the sufficient condition becomes
\begin{align}
\gD \xi_{n+1,2} + \gD \xi_{n+1,n-1} &\geq 0 \text{ (downscale)}, \\
\gD \xi_{n+1,2} + \gD \xi_{n+1,n-1} &\leq 0 \text{ (upscale). } 
\end{align}
We see that locality survives even the violation of the fusion rule hypothesis if $\gD\xi_{n+1,2}$ and $\gD\xi_{n+1,n-1}$ are both  positive downscale and negative upscale.

\section{Stability of the upscale and downscale cascade}

We now turn to the question of statistical stability with respect to forcing perturbations.  Statistical stability is defined as the requirement that there should be a region $\cJ_n$ such that $Q_n (\{\bfX\}_n )$ is negligible relative to contributions to $D_{kn}(\{\bfX\}_n )$ for all $\{\bfX\}_n \in \cJ_n$ in that region. Even when the forcing spectrum is confined to a narrow range of scales, it is not self-evident that  this requirement is satisfied, due to feedback loops of $F_n$ onto $Q_n$ (see below). 

The first explicit proof that the inertial range of three-dimensional turbulence is statistically stable was given by L'vov and  Procaccia in section II-C-3 of \cite{article:Procaccia:1996}.  The proof used the balance equations of the standard structure functions (not the generalized structure functions used in this paper), and it covers the case of stability with respect to gaussian forcing when the scaling exponents $\gz_n$ take Kolmogorov scaling values $\gz_n =n/3$. The value of this proof has gone by unnoticed   because experiments and numerical simulations have established the statistical stability of the three-dimensional energy range beyond all doubt.  For the problem of two-dimensional turbulence however, where the lack of robustness of the upscale and downscale cascades is the unresolved  problem,  the method used by L'vov and Procaccia in that proof  is very illuminating. The main idea is  to estimate the scaling exponent of the ratio $Q_n/D_{kn}$ and require the appropriate constraint on that exponent such that the ratio vanishes asymptotically in the inertial range, in the limit of extending the range. 

Consider a geometry of velocity differences $\{\bfx\}_n$ such that all point to point distances have order of magnitude $1$, and define the scaling exponent $q_n$ by
\begin{equation}
Q_{n}(R) \equiv Q_{n}(R\{\bfx\}_n)  \sim \fracp{R}{\ell_0}^{q_n}.
\end{equation}
with $R$ a scale in the inertial range. From locality (proved in the previous section) we also know that, 
\begin{equation}
D_{kn} (R) \equiv D_{kn} (R\{\bfx\}_n)\sim   \fracp{R}{\ell_0}^{\gz_{n+1}-1}.
\end{equation}
It follows that the ratio $Q_n/D_{kn}$ scales as
\begin{equation}
\frac{Q_{n}(R)}{D_{kn} (R)} \sim \fracp{R}{\ell_0}^{q_n-( \gz_{n+1}-1)}.
\end{equation}
In a direct cascade, such as the energy cascade of three-dimensional turbulence and the enstrophy cascade of two-dimensional turbulence, this ratio must vanish in the limit $\ell_0 \rightarrow +\infty$.  It follows that the condition for the statistical stability of a downscale cascade reads 
\begin{equation}
\gD q_n \equiv q_n- (\gz_{n+1}-1)>0,\; \forall n\in\bbN, n>1.
\end{equation}
In an upscale cascade, such as the inverse energy cascade of two-dimensional turbulence, the same ratio must vanish in the limit $\ell_0 \rightarrow 0$.  This leads to the same condition with the inequality reversed: 
\begin{equation}
\gD q_n \equiv  q_n- (\gz_{n+1}-1) < 0,\; \forall n\in\bbN, n>1.
\end{equation}

\subsection{The case of gaussian forcing}

For the simplest case of Gaussian delta-correlated in time forcing, the exponents $q_n$ can be calculated  in terms of $\gz_n$. This makes it possible to investigate statistical stability rigorously. 

We begin with the assumption that $f_{\ga}$ is a delta-correlated stationary gaussian field with $\avg{f_{\ga}(\bfx)}=0$, and 
\begin{equation}
\avg{f_{\ga}(\bfx_1, t_1)f_{\gb}(\bfx_2, t_2)} = 2\gee C_{\ga\gb}(\bfx_1,\bfx_2)\gd (t_1-t_2),
\end{equation}
where $\gee$ is constant, and $C_{\ga\gb}$ is normalized such that $C_{\ga\ga}(\bfx,\bfx) =1$.   Without loss of generality we may assume that $\pd_\ga f_\ga =0$, and therefore $\cP_{\ga\gb} f_\gb = f_\gb$. Thus, we have the identity
\begin{equation}
\int d\bfy\; P_{\gb\gc}(\bfx_2 - \bfy) C_{\ga\gc}(\bfx_1, \bfy) = C_{\ga\gb}(\bfx_1, \bfx_2), 
\end{equation}
which will be used below.

We define the forcing scale $\ell_0$ from the Taylor expansion
\begin{equation}
C_{\ga\gb}(\bfx+\bfy,\bfx) = \frac{\gd_{\ga\gb}}{d} - A_{\ga\gb}^{(2)} \fracp{\nrm{\bfy}}{\ell_0}^2 + O(\ell_0^{-4}),
\end{equation}
valid in the limit $\nrm{\bfy} \ll \ell_0$. Note that the odd-order terms vanish by incremental isotropy. In the limit  $\nrm{\bfy} \gg \ell_0$, on the other hand, we have the asymptotic expansion
\begin{equation}
C_{\ga\gb}(\bfx+\bfy,\bfx) \sim \fracp{\ell_0}{\nrm{\bfy}}^a \left[ A_{\ga\gb}^{(0)} + A_{\ga\gb}^{(1)}\fracp{\ell_0}{\nrm{\bfy}} + O(\ell_0^2) \right].
\end{equation}
Note that $a$, which is  an unspecified scaling exponent dependent on our choice of stochastic forcing, must satisfy $a>0$,  since the correlation must vanish at $\nrm{\bfy} \goto +\infty$.  Also note that $\gee$ is the total rate of energy injection. In general, the work done on the fluid  is $\gee_{in} (\bfx) = f_{\ga}(\bfx) u_{\ga}(\bfx)$. For delta-correlated forcing, it is easy to show that $\avg{\gee_{in} (\bfx)} = \gee C_{\ga\ga} (\bfx, \bfx)$ (see proof in appendix \ref{sec:onetimeresponse}).

Recall that the total forcing term $Q_n$ is given by 
\begin{equation}
Q_n(\{\bfX\}_n,t) = \sum_{k=1}^n Q_{kn}(\{\bfX\}_n^k,\bfX_k, t),
\end{equation}
where $Q_{kn}$ reads
\begin{widetext}
\begin{equation}
Q_{kn}^{\ga_1\ga_2\cdots\ga_{n-1}\gb}(\{\bfX\}_{n-1},\bfY, t) = \avg{\left[ \prod_{k=1}^{n-1} w_{\ga_k} (\bfX_k, t) \right] \gf_{\gb}(\bfY, t)},
\end{equation}
with $\gf_{\ga}(\bfX, t) = f_{\ga}(\bfx, t) - f_{\ga}(\bfxp, t)$. For Gaussian forcing, it can be shown (see appendix \ref{sec:gaussianforcing}) that the forcing contributions $Q_{kn}$ to the generalized balance equations read
\begin{equation}
Q_{kn}^{\ga_1\cdots\ga_{n-1}\gb} (\{\bfX\}_{n-1}, \bfY,t) = \sum_{l=1}^{n-1} F_{n-2}^{\ga_1\cdots\ga_{l-1}\ga_{l+1}\cdots\ga_{n-1}}(\{\bfX\}_{n-1}^l) Q_{\ga_l\gb} (\bfX_l, \bfY),
\label{eq:gforceA}
\end{equation}
with $Q_{\ga\gb}(\bfX,\bfY)$ given by
\begin{align}
Q_{\ga\gb}(\bfX,\bfY) &= \avg{w_\ga (\bfX,t) \gf_\gb (\bfY,t)} =  2\gee\int d\bfz \; [P_{\ga\gc}(\bfx-\bfz) - P_{\ga\gc}(\bfxp-\bfz)][C_{\gb\gc} (\bfy,\bfz) - C_{\gb\gc} (\bfyp,\bfz)]\\
&= 2\gee [C_{\ga\gb} (\bfy, \bfx) - C_{\ga\gb} (\bfyp, \bfx) - C_{\ga\gb} (\bfy, \bfxp) + C_{\ga\gb} (\bfyp, \bfxp)].
\label{eq:gforceB}
\end{align}
\end{widetext}
 The physical intuition is that there is a feedback loop between forcing, whose spectrum is defined by $Q_{\ga\gb}(\bfX,\bfY)$, and the resulting behavior of turbulence which is captured by the structure functions $F_n$. More specifically, we see that $F_{n-2}$ provides feedback to $Q_n$, when the forcing is gaussian. For statistical stability we need this feedback to be negligible in the inertial range.

The immediate implication of eq. \eqref{eq:gforceA} is that $q_n = \gz_{n-2} + q_2$ with $q_2$ the scaling exponent of $Q_{\ga\gb}$. It follows that 
\begin{equation} 
\gD q_n = (\gz_{n-2} + q_2)-(\gz_{n+1} - 1).
\end{equation}
The remaining challenge is to calculate $q_2$. We will see that $q_2$ depends on whether the cascade is upscale or downscale. In the rest of this section, we will derive the separate stability conditions for a downscale cascade and for an upscale cascade. 

\subsection{Stability conditions for downscale cascades}

For the case of a downscale cascade, using the Taylor expansion of $Q_{\ga\gb}(\bfX,\bfY)$  in the limit $\nrm{\bfX-\bfY} \goto \bfzero$, the scaling of $Q_{\ga\gb}$ can be estimated as
\begin{widetext}
\begin{align}
Q_{\ga\gb}(\bfX,\bfY) &= 2\gee [C_{\ga\gb} (\bfy, \bfx) - C_{\ga\gb} (\bfyp, \bfx) - C_{\ga\gb} (\bfy, \bfxp) + C_{\ga\gb} (\bfyp, \bfxp)]\\
 &= 2\gee [(C_{\ga\gb} (\bfy, \bfx)- C_{\ga\gb} (\bfx, \bfx))-(C_{\ga\gb} (\bfyp, \bfx)- C_{\ga\gb} (\bfx, \bfx))-(C_{\ga\gb} (\bfy, \bfxp)- C_{\ga\gb} (\bfxp, \bfxp))\\
&\quad +(C_{\ga\gb} (\bfyp, \bfxp)- C_{\ga\gb} (\bfxp, \bfxp))]\\
 &\sim (2\gee/\ell_0^2) [\nrm{\bfy - \bfx}^2-\nrm{\bfyp - \bfx}^2-\nrm{\bfy - \bfxp}^2+\nrm{\bfyp - \bfxp}^2]\sim \gee (R/\ell_0)^2,
\end{align}
\end{widetext}
which suggests that for a downscale cascade,  $q_2 = 2$. It is easy to see that for a monofractal velocity field with $\gz_n=nh$, the stability condition reads 
\begin{align}
\gD q_n &=  (\gz_{n-2} + 2)-(\gz_{n+1} - 1)\\
&= 3-3h > 0, \forall n\in\bbN : n>1,
\end{align}
which requires $h<1$. In a multifractal case one has a linear combination of independent monofractal contributions, and it can be shown that the constraint $0 < \gz_3 < 3$ is a sufficient condition for statistical stability. This follows from the inequality $\gz_{n+1}\leq\gz_3+\gz_{n-2}$ (see appendix~\ref{sec:holder}):
\begin{align}
\gD q_n &=\gz_{n-2}-\gz_{n+1}+3 \\
&\ge\gz_{n-2}-\gz_{n-2} -\gz_3 +3 \\
&= 3-\gz_3>0, \forall n\in\bbN : n>2.
\end{align}
For $n=2$, we get $\gD q_2 = q_2 - (\gz_3 - 1) = 3-\gz_3$, which implies, from the stability condition $\gD q_2>0$,  that $0 < \gz_3 < 3$ is also a \emph{necessary} condition. 

For the case of the downscale energy cascade of three-dimensional turbulence we have $\gz_3=1$ , which can be  derived from the solvability condition for the homogeneous equation $\cO_2 F_3=0$ \cite{article:Procaccia:1996:2,article:Procaccia:1996:3,article:Gkioulekas:2007}.  This satisfies the sufficient condition $0 < \gz_3 < 3$ for statistical stability  very generously, so it is hardly a surprise that the energy cascade is so  robust.  Also worth noting is that for a hypothetical downscale helicity cascade we have $\gz_3=2$ , which also satisfies the stability condition. 

The story changes for the case of the downscale enstrophy cascade.  We know, from combining the Eyink and Falkovich-Lebedev  theories of the two-dimensional  enstrophy cascade \cite{article:Lebedev:1994,article:Lebedev:1994:1,article:Eyink:2001,article:Tung:2006}, that when it exists with constant enstrophy flux, the enstrophy cascade has no intermittency corrections.  Thus, the scaling exponents $\gz_n$ all satisfy the monofractal scaling $\gz_n=n$ , which implies that 
\begin{equation}
\gD q_n =\gz_{n-2}-\gz_{n+1}+3=0.
\end{equation}
So,  we have a borderline situation where the stability condition is neither satisfied nor broken! Consequently, the actual stability of the downscale enstrophy cascade is not decided by scaling exponents but by the numerical coefficients in front of $Q_n$ and $D_{kn}$ .  This is where it gets interesting.

The leading contribution to $Q_n$ is proportional to the total rate of energy injection $\gee$.   However, one should bear in mind that  the downscale enstrophy cascade is  forced by the combined effect of both the forcing term $f_\ga$ and the large-scale dissipation term $(-1)^{m+1}\gb \del^{-2m}u_{\ga}$.   As a result of this combined forcing, the enstrophy cascade is injected with a smaller  enstrophy rate $\gn_{uv}$ and a very small energy rate $\gee_{uv}$ with $\gn_{uv} <\gn$ and $\gee_{uv} \ll\gee$. If we assume that this combined effect itself can be modelled as gaussian forcing, then the leading contribution to the effective forcing on the enstrophy cascade is proportional only to the rate $\gee_{uv}$ of the subleading downscale energy flux.  Because $\gee_{uv}$ vanishes rapidly as the separation of scales in the enstrophy cascade is increased \cite{article:Eyink:1996,article:Tung:2005:1}, this leading contribution can be made as small as desired simply by taking the limit $\nu \goto 0^+$. For small enough downscale energy flux $\gee_{uv}$, the next order term with $q_2 \geq 3$ becomes dominant, and combined with $\gz_n=n$ it is easy to show that the stability condition is now $\gD q_n >0$. 

The conclusion from this analysis is that the stability of the downscale enstrophy cascade requires that the accompanying downscale energy flux should be very small.  For that to happen, we need two things: First,  it is necessary to have a dissipation sink at large scales to absorb most of the injected energy at the forcing scale or at larger scales. Second, we must   have a large separation of scales between the forcing scale and the dissipation scale at small scales, which means that a significant amount of numerical resolution is required.  These two requirements, we believe, are the reason why it has been so difficult to reproduce the enstrophy cascade in numerical simulations.  It is worth noting that Tran and Bowman \cite{article:Bowman:2004} came to a similar conclusion by a different argument, that the robustness of the downscale enstrophy cascade requires a vanishing downscale energy flux.

\subsection{Stability conditions for upscale cascades}

The fundamental difference between an upscale  cascade and a downscale  cascade with respect to stability is that in the  upscale cascade   the \Holder inequalities now take the form $\gz_{n+k} \geq \gz_n + \gz_k$, and  the condition for statistical stability reads $\gD q_n <0 ,\;\forall n\ge 1$.  We will now prove that inverse cascades are always statistically stable with respect to variations in the forcing statistics, provided that $\gz_3 \geq 1$.  This is consistent with the numerical evidence that the inverse energy cascade is much easier to obtain in simulations than the direct enstrophy cascade.

Again, using Taylor expansion in the limit $\nrm{\bfX-\bfY}\goto\infty$, we see that $Q_{\ga\gb}$ scales as
\begin{widetext}
\begin{align}
Q_{\ga\gb}(\bfX,\bfY) &= 2\gee [C_{\ga\gb} (\bfy, \bfx) - C_{\ga\gb} (\bfyp, \bfx) - C_{\ga\gb} (\bfy, \bfxp) + C_{\ga\gb} (\bfyp, \bfxp)]\\
&\sim  2\gee \left[ \fracp{\ell_0}{\nrm{\bfy - \bfx}}^a - \fracp{\ell_0}{\nrm{\bfyp - \bfx}}^a - \fracp{\ell_0}{\nrm{\bfy - \bfxp}}^a + \fracp{\ell_0}{\nrm{\bfyp - \bfxp}}^a \right], 
\end{align}
\end{widetext}
which gives $q_2 = -a <0$. For a monofractal velocity field with $\gz_n=nh$, the stability condition reads
\begin{align}
\gD q_n &= q_n - (\gz_{n+1}-1)\\
&= q_2+\gz_{n-2} - (\gz_{n+1}-1) \\
&= q_2+1-3h < 0, \forall n\in\bbN : n>1.
\end{align}
Since $q_2 <0$, the condition $h\geq 1/3$ is sufficient. For the more general multifractal case, using the inequality $\gz_{n+1}\geq\gz_3+\gz_{n-2}$, we can upper-bound $\gD q_n$ as follows:
\begin{align}
\gD q_n &= q_2+\gz_{n-2} - (\gz_{n+1}-1) \\
&\leq q_2+1-(\gz_{n-2}+\gz_3) + \gz_{n-2} \\
&= q_2 + 1 - \gz_3, \forall n\in\bbN : n>2.
\end{align}
For $n=2$, we get an equality:  $\gD q_2 = q_2 - (\gz_3 - 1)$. Thus, the stability condition $\gD q_n < 0$ is satisfied when $\gz_3 > q_2+1$, which is indeed satisfied when $q_2 <0$ and $\gz_3 \geq 1$. For the inverse energy cascade of two-dimensional turbulence, we have $\gz_3 =1$ which satisfies the requirements for stability. 

There is however another effect that can  destabilize the inverse energy cascade. We have shown in a previous paper \cite{article:Gkioulekas:2007} that  the loss of asymptotic  homogeneity by  the effect of the boundary conditions on the flow  amplifies the sweeping term $I_n$ at  the large scales.  As a result,  at sufficiently large length scales, the ratio  $I_n/D_{kn}$ becomes significant, and excites a particular solution superimposed on top of the homogeneous solution associated with the inverse cascade. The particular solution   corresponds to the coherent structures associated with the ``energy condensation effect''.  The formation of these  coherent structures  is very likely to further intensify the ratio $I_n/D_{kn}$.  As we have explained in the introduction, it has been shown that if these coherent vortices are removed before the evaluation of the energy spectrum, the usual inverse energy cascade spectrum is recovered \cite{article:Borue:1994,article:Gurarie:2001:1,article:Fischer:2005}. This result is consistent with our theory, and it confirms  that  the homogeneous solution, corresponding to the inverse energy cascade, exists side by side with the particular solution, corresponding to the coherent structures, even when the particular solution is dominant.  The possible role of the sweeping term on the stability of the enstrophy cascade is currently not well-understood.

\section{Conclusion and Discussion}
\label{sec:conclusion}

We have shown that the non-perturbative locality of the inertial ranges of two-dimensional turbulence is an immediate consequence of the fusion rules hypothesis. The physical interpretation of what we have done is to prove, strictly in the context of the incompressible Navier-Stokes equation, that \emph{universality implies locality}. A proof of the fusion rules by diagrammatic theory is essentially the converse and more interesting claim: that \emph{locality implies universality}.  This result leads to an apparently curious paradox: the usual   understanding of locality, in terms of triad interactions in Fourier space, suggests that a necessary condition for locality is that the energy spectrum $E(k)$ must have slope between $k^{-3}$ and $k^{-1}$.  This corresponds to the inequality $0<\gz_2<2$.   The paradox is that this constraint does not appear anywhere in our locality proof! In recent work, Eyink \cite{article:Eyink:2005} investigated the locality of the downscale enstrophy cascade and the inverse energy cascade using a filtering method \cite{article:Germano:1992,article:Eyink:1995:1,article:Eyink:1995:2}.  His argument also leads to the inequality $0<\gz_2<2$ as a  sufficient locality condition. It follows that whereas the inverse energy cascade is local the direct enstrophy cascade is IR marginally-nonlocal. Unlike the argument in this paper, Eyink's argument has only considered the kinematic locality of the flux term  and not the statistical locality associated with unfused higher-order structure functions. On the other hand, our argument is less rigorous in its present form, as it assumes the fusion rules without proof.

A fundamental problem with establishing  locality in   Fourier space is that the Fourier transform involves an integral that ranges over every length scale, including the forcing  length scales and the dissipation length scales.  To preserve locality, the main contribution to the integral must come from the inertial range. The inequality $0<\gz_2<2$ comes in as a  necessary condition for the survival of locality under the Fourier integral \cite{book:Frisch:1995}. The same issue arises when locality is characterized with a filtering transform (i.e. forward Fourier, truncation, backward Fourier), as was done by Eyink \cite{article:Eyink:2005}, albeit with  a broader definition of filtering.  Beyond that, the underlying argument based on diagrammatic theory \cite{article:Procaccia:1995:1,article:Procaccia:1995:2,article:Procaccia:1996} that justifies the fusion rules hypothesis itself  can  impose further constraints on $\gz_2$, which still need to be investigated carefully. For example, one other way the constraint $0<\gz_2<2$ can come in is if we require perturbative locality  for each Feynman diagram \cite{article:Procaccia:1995:1}. Perturbative locality may be a necessary condition for the fusion rules hypothesis. If that is true, then perhaps $0<\gz_2 <2$ is implicitly assumed when we postulate the fusion rules hypothesis. We have also shown in this paper that the related condition $0<\gz_3 <3$ is required for stability under Gaussian forcing, which is as essential as locality for the existence of a universal inertial range.

It should be stressed  that  any  constraints  on scaling exponents needed only to prove the fusion rules hypothesis by Feynman diagrams , are needed only to establish the \emph{universality} of the scaling exponents $\gz_n$ of the inertial range. We \emph{should} expect to find that the conditions for locality itself are weaker.  For example, we have shown in this paper that locality is possible even when the fusion rules fail, provided that the fusion exponents $\xi_{np}$ deviate in the correct direction. In fact, it is possible to have local interactions, as per our definition, even when the underlying diagrammatic theory does not yield local Feynman diagrams! This scenario is not entirely hypothetical; in the case studied by Ref.\onlinecite{article:Guzdar:2000} of an  enstrophy range under strong Ekman  dissipation, this may be  precisely what happens, with the slope being steeper than $k^{-3}$ and non-universal, but still allowing an appearently local enstrophy cascade to exist. 

The key idea that can help us unravel these paradoxes is that the non-perturbative locality studied in this paper is a weaker condition than perturbative locality.  Non-perturbative locality requires only the combined effect of all Feynman diagrams to be local.  Perturbative locality, on the other hand, requires that   each diagram individually should be local.  This distinction between perturbative and  non-perturbative locality may clarify the paradoxical situation with the enstrophy cascade where the spectrum of the enstrophy cascade is  consistent with a dimensional analysis argument based on a locality assumption  even though the slope is too steep to be self-consistent with that assumption! Adding a logarithmic correction resolves the situation in a one-loop closure model \cite{article:Kraichnan:1971:2}, and the combination of more recent results by Falkovich and Lebedev \cite{article:Lebedev:1994:1} and Eyink \cite{article:Eyink:2001} suggest that  the same logarithmic correction persists for the exact theory, with no higher-order adjustments.  Nevertheless, a reconcilliation of the spectrum slope and the locality requirement is still an ``uncomfortable'' notion, to say the least. We believe that a possible resolution of this paradox  is to claim that the  enstrophy cascade is local in the  non-perturbative sense, as far as the exact theory is concerned,   and borderline non-local only in the perturbative sense. From a physical standpoint  the relevant locality needed as  a precondition for establishing the existence of an inertial range is the non-perturbative locality. However,  some confusion can arise from the fact that closure models unwittingly exchange  non-perturbative locality with perturbative locality!

The careful reader will note that the   non-perturbative locality  is  also weaker  than the more intuitive (and less rigorous) physical understanding   of  locality as the notion that  the effect of the forcing range and dissipation range is ``forgotten'' in the inertial range. We may designate locality, in this sense, as ``strong'' locality, so that it can be distinguished from the weaker  non-perturbative statistical locality. The proposed theory can help make the meaning of  this notion of ``strong'' locality more rigorous.  The key idea is that  it is possible to have local interaction integrals in the contributions to the $\cO_n F_{n+1}$ term of the balance equations and still pick up an effect from the forcing range or the dissipation range into the multi-dimensional regions $\cJ_n$ that  are supposed to be the inertial range, in our generalized sense. It all  depends on how much forcing and dissipation ``wish to creep into'' the inertial range.  We can find that out by comparing the magnitude of the  $Q_n$ , $I_n$, and $\cD_n F_n$ terms of the generalized balance equations against    the magnitude of the contributions $D_{kn}$ to the interaction term. Thus,   we find that there are three distinct conditions that need to hold to have strong locality:  first, the interaction integral itself has to be local;  second, we need to establish the property of \emph{statistical stability} which will guarantee that  the forcing effect $Q_n$ and  the sweeping interactions $I_n$ do not creep into the inertial range;  third, a calculation of the shape of the dissipation range can show whether there is a wide enough region $\cJ_n$ in which the  dissipation term $\cD_n F_n$ is negligible. One  advantage of the generalized balance equations framework is that it allows us to account mathematically for these three distinct effects separately.

In this paper, we examined only the first condition and part of the second condition. We have shown that statistical stability with respect to forcing applies unconditionally for the inverse energy cascade. For the enstrophy cascade, statistical stability requires large-scale dissipation and a vanishing downscale energy dissipation. For any downscale cascade in general, stability constrains the corresponding \Holder exponent as $h<1$. For an upscale cascade,  the corresponding constraint is $h\geq 1/3$. We began considering the role of sweeping  in a previous paper \cite{article:Gkioulekas:2007}, and the role of the dissipation term will be studied in future work.

\begin{acknowledgements}
It is a pleasure to thank Ka-Kit Tung for his advice and encouragement. The research is supported  in part by the National Science Foundation, under grant DMS-03-27658.
 \end{acknowledgements}

\appendix

\section{Derivation of the balance equations}
\label{app:balanceproof}

In this appendix we give a detailed derivation of the generalized balance equations. Recall that we defined the generalized structure function $F_n$ as 
\begin{equation}
F_n(\{\bfX \}_n, t) = \left\langle \left [  \prod_{k=1}^n w_{\ga_k} (\bfX_{k} , t) \right] \right\rangle.   
\end{equation} 
\begin{widetext}
By differentiating $F_n$ with respect to $t$ and substituting the Navier-Stokes equations we obtain:
\begin{align}
\pderiv{F_n (t)}{t} &= \sum_{k=1}^n \avg{\pderiv{w_{\ga_k}(\bfx_k, \bfxp_k,t)}{t}\left[ \prod_{l=1,l\neq k}^n w_{\ga_l}(\bfx_l, \bfxp_l,t) \right]}
= \sum_{k=1}^n [ -N_{kn} + Q_{kn} ] + \nu J_n + \gb H_n. 
\end{align}

Here, the terms $\nu J_n$ and $\gb H_n$ are the contributions of the small-scale and large-scale sinks with
\begin{align}
 J_n^{\ga_1\ga_2\cdots\ga_n}  (\{\bfx, \bfxp\}_n , t) &=  \sum_{k=1}^n (\del_{\bfx_k}^{2\gk} + \del_{\bfxp_k}^{2\gk}) F_n (\{\bfx, \bfxp\}_n , t),\\
 H_n^{\ga_1\ga_2\cdots\ga_n}  (\{\bfx, \bfxp\}_n , t) &= \sum_{k=1}^n (\del_{\bfx_k}^{-2m} + \del_{\bfxp_k}^{-2m}) F_n (\{\bfx, \bfxp\}_n , t),
\end{align}
 where $\lapl_{\bfx_k}$ is the Laplacian with respect to $\bfx_k$; $\lapl_{\bfxp_k}$ is the Laplacian with respect to $\bfxp_k$. Also,  $N_{kn}$ represents the contributions of $\cP_{\ga\gb}\pdc (u_{\gb} u_{\gc}) $, and $Q_{kn}$ represents the contributions of $\cP_{\ga\gb} f_{\gb}$, and they  read:
\begin{align}
 Q_{kn}^{\ga_1\ga_2\cdots\ga_n} (\{\bfx, \bfxp\}_n , t) &= \avg{\left[ \prod_{l=1,l\neq k}^n w_{\ga_l}(\bfx_l, \bfxp_l,t) \right]\cP_{\ga_k\gb}(f_{\gb}(\bfx_k,t) - f_{\gb}(\bfxp_k,t))}, \\
 N_{kn}^{\ga_1\ga_2\cdots\ga_n} (\{\bfx, \bfxp\}_n , t) &= \avg{\left[ \prod_{l=1,l\neq k}^n w_{\ga_l}(\bfx_l, \bfxp_l,t) \right]\cP_{\ga_k\gb}[\pd_{\gc,\bfx_k}(u_{\gb,\bfx_k} u_{\gc,\bfx_k}) - \pd_{\gc,\bfxp_k}(u_{\gb,\bfxp_k} u_{\gc,\bfxp_k})]} \\
&=  \avg{\left[ \prod_{l=1,l\neq k}^n w_{\ga_l}(\bfx_l, \bfxp_l,t) \right]\cP_{\ga_k\gb} \cN_{\gb} (\bfx_k, \bfxp_k,t)}\\
&= \avg{\left[ \prod_{l=1,l\neq k}^n w_{\ga_l}(\bfx_l, \bfxp_l,t) \right]\int d\bfy P_{\ga_k\gb} (\bfy) \cN_{\gb} (\bfx_k-\bfy, \bfxp_k-\bfy,t)}.
\end{align}

  Here we use the abbreviations $u_{\ga,\bfx_k} = u_{\ga} (\bfx_k, t)$ and $u_{\ga,\bfxp_k} = u_{\ga} (\bfxp_k, t)$,  $w_{\ga, k} = w_{\ga}(\bfx_k, \bfxp_k,t)$, and $\pd_{\ga,\bfx_k}$ is the spatial derivative in the $\ga$ direction with respect to $\bfx_k$. Also, $\cN_{\gb} (\bfx_k, \bfxp_k,t)$ is  the non-linear factor defined as:
\begin{align*}
 \cN_{\gb} (\bfx_k, \bfxp_k,t) &= \pd_{\gc,\bfx_k}(u_{\gb,\bfx_k} u_{\gc,\bfx_k}) - \pd_{\gc,\bfxp_k}(u_{\gb,\bfxp_k} u_{\gc,\bfxp_k}) = u_{\gc,\bfx_k}\pd_{\gc,\bfx_k}(u_{\gb,\bfx_k}-u_{\gb,\bfxp_k})  + u_{\gc,\bfxp_k}\pd_{\gc,\bfxp_k}(u_{\gb,\bfx_k}- u_{\gb,\bfxp_k}) \\
 &= \pd_{\gc,\bfx_k} (u_{\gc,\bfx_k} w_{\gb,k}) +  \pd_{\gc,\bfxp_k} (u_{\gc,\bfxp_k} w_{\gb,k})= u_{\gc,\bfx_k} \pd_{\gc,\bfx_k} w_{\gb,k} +  u_{\gc,\bfxp_k}\pd_{\gc,\bfxp_k} w_{\gb,k}.
\end{align*}
\end{widetext}


It is easy to see that the nonlinear terms $N_{kn} $ cannot be written exclusively in terms of velocity differences. The remarkable characteristic of the derivation of the balance equations by  L'vov and  Procaccia \cite{article:Procaccia:1996:3} is that the nonlinear term $N_{kn}$ is rearranged as the sum of a local term  $D_{kn}$ and a sweeping term  $I_{kn}$ such that the local term can be expressed as a linear operator on $F_{n+1}$. Although L'vov and  Procaccia \cite{article:Procaccia:1996:3}  eliminated the sweeping term on the grounds of global homogeneity, we believe it is appropriate to retain it here in its simplified form. 

To isolate the sweeping term, we define a generalized mean velocity  $\cU_{\ga} (\{\bfz, \bfzp\}_n , t)$ as:
  \begin{equation}
  \cU_{\ga} (\{\bfz, \bfzp\}_n , t) = \frac{1}{2n} \sum_{k=1}^n (u_{\ga} (\bfz_k, t) + u_{\ga} (\bfzp_k, t)),
  \end{equation}
\begin{widetext}
  and the corresponding velocity fluctuation
\begin{align}
 v_{\ga} (\bfx, \{ \bfz, \bfzp\}_n, t) &= u_{\ga} (\bfx, t) - \cU_{\ga} (\{\bfz, \bfzp\}_n , t) = \frac{1}{2n} \sum_{k=1}^n [w_{\ga} (\bfx, \bfz_k) + w_{\ga} (\bfx, \bfzp_k)].
\end{align}

 We may then decompose $ \cN_{\ga} (\bfx_k, \bfxp_k,t)$, in general, to
\begin{equation}
\cN_{\ga} (\bfx_k, \bfxp_k,t) = \cS_{\ga}(\bfx_k, \bfxp_k, \{ \bfz, \bfzp\}_n,t) + \cL_{\ga}(\bfx_k, \bfxp_k, \{ \bfz, \bfzp\}_n,t),
\end{equation}
 where $\cS_{\ga}$ and $\cL_{\ga}$  are defined as:
 \begin{equation}
 \begin{split}
\cS_{\ga}(\bfx_k, \bfxp_k, \{ \bfz, \bfzp\}_n,t) &= \cU_{\gb} (\{ \bfz, \bfzp\}_n , t)  (\pd_{\gb, \bfx_k} + \pd_{\gb, \bfxp_k}) w_{\ga} (\bfx_k, \bfxp_k, t),\\
\cL_{\ga}(\bfx_k, \bfxp_k, \{ \bfz, \bfzp\}_n,t) &= [v_{\gb} (\bfx_k, \{ \bfz, \bfzp\}_n, t)\pd_{\gb, \bfx_k} + v_{\gb} (\bfxp_k, \{ \bfz, \bfzp\}_n, t)\pd_{\gb, \bfxp_k}]  w_{\ga} (\bfx_k, \bfxp_k, t).
 \end{split}
 \end{equation}
In general, $\{ \bfz, \bfzp\}_n$ can be chosen any way we wish. Here, we specifically use the choice:
\begin{equation}
\cN_{\gb} (\bfx_k, \bfxp_k,t) = \cL_{\gb}(\bfx_k, \bfxp_k, \{ \bfx, \bfxp\}_n, t) + \cS_{\gb}(\bfx_k, \bfxp_k, \{ \bfx, \bfxp\}_n, t).
\end{equation}
This gives the decomposition $N_{kn} = D_{kn} + I_{kn}$ with
\begin{align}
 D_{kn}^{\ga_1\ga_2\cdots\ga_n} (\{\bfx, \bfxp\}_n , t) &=  \avg{\left[ \prod_{l=1,l\neq k}^n w_{\ga_l}(\bfx_l, \bfxp_l,t) \right]\int d\bfy P_{\ga_k\gb} (\bfy)\cL_{\gb}(\bfx_k-\bfy, \bfxp_k-\bfy, \{ \bfx, \bfxp\}_n, t)}, \label{eq:dkn} \\
I_{kn}^{\ga_1\ga_2\cdots\ga_n} (\{\bfx, \bfxp\}_n , t) &= \avg{\left[ \prod_{l=1,l\neq k}^n w_{\ga_l}(\bfx_l, \bfxp_l,t) \right]\int d\bfy P_{\ga_k\gb} (\bfy) \cS_{\gb}(\bfx_k-\bfy, \bfxp_k-\bfy, \{ \bfx, \bfxp\}_n, t)}. 
\end{align}
Here $I_{kn}$ represents the sweeping interactions and $D_{kn}$ represents the local interactions.


The sweeping term $I_{kn}$ can be simplified as follows: We use the decomposition $P_{\ga\gb} (\bfx) = \gd_{\ga\gb} \gd (\bfx)  - P_{\ga\gb}^{\parallel} (\bfx)$ to split $I_{kn}$ to two terms: $I_{kn} = I_{kn}^{(1)} + I_{kn}^{(2)}$ with $I_{kn}^{(1)}$ corresponding to $\gd_{\ga\gb} \gd (\bfx)$ and  $I_{kn}^{(2)}$ corresponding to $P_{\ga\gb}^{\parallel} (\bfx)$. We also use $\cP_{\ga\gb}^{\parallel} u_{\gb} = 0$. The integral inside the ensemble average of $I_{kn}$  splits to two parts: $I_1$ and $I_2$.  The first part  $I_1$ reads:
\begin{align}
I_1 &= \int d\bfy \; \gd_{\ga_k\gb} \gd (\bfy)\cS_{\gb}(\bfx_k-\bfy, \bfxp_k-\bfy, \{\bfx,\bfxp\}_n,t)= \cS_{\ga_k}(\bfx_k, \bfxp_k, \{\bfx,\bfxp\}_n,t)\\
&= \cU_{\gc}(\{\bfx,\bfxp\}_n,t) (\pd_{\gc,\bfx_k} + \pd_{\gc,\bfxp_k})w_{\ga_k} (\bfx_k, \bfxp_k, t).
\end{align}
The second part $I_2$ is shown to be zero by incompressibility:
\begin{align}
 I_2 &= \int d\bfy P_{\ga\gb}^{\parallel} (\bfy) \cS_{\gb}(\bfx_k-\bfy, \bfxp_k-\bfy,\{\bfx,\bfxp\}_n, t) \\
&=  \int d\bfy P_{\ga\gb}^{\parallel} (\bfy)  \cU_{\gc}(\{\bfx,\bfxp\}_n,t) (\pd_{\gc,\bfx_k} + \pd_{\gc,\bfxp_k})w_{\gb} (\bfx_k-\bfy, \bfxp_k-\bfy, t) \\
&= \cU_{\gc}(\{\bfx,\bfxp\}_n,t) (\pd_{\gc,\bfx_k} + \pd_{\gc,\bfxp_k}) \int d\bfy P_{\ga\gb}^{\parallel} (\bfy)  w_{\gb} (\bfx_k-\bfy, \bfxp_k-\bfy, t) = 0.
\end{align}
Because $P_{\ga\gb}^{\parallel}$ is the nonlocal part of the projection operator, this result implies that the pressure effect does not contribute to the sweeping interactions or to the violation of incremental homogeneity. Thus,  $I_{kn}$ is determined by  $I_1$  and it simplifies to
\begin{align}
 I_{kn}^{\ga_1\ga_2\cdots\ga_n} (\{\bfx, \bfxp\}_n , t)&= \avg{\left[ \prod_{l=1,l\neq k}^n w_{\ga_l}(\bfx_l, \bfxp_l,t) \right]\cU_{\gc}(\{\bfx,\bfxp\}_n,t) (\pd_{\gc,\bfx_k} + \pd_{\gc,\bfxp_k})w_{\ga_k} (\bfx_k, \bfxp_k, t) } \\
&=  (\pd_{\gc,\bfx_k} + \pd_{\gc,\bfxp_k}) \avg{\cU_{\gc}(\{\bfx,\bfxp\}_n,t)\left[ \prod_{l=1}^n w_{\ga_l}(\bfx_l, \bfxp_l,t) \right]}.
\end{align}
This result was given previously by L'vov and Procaccia in section IV-B and appendix B of Ref.\onlinecite{article:Procaccia:1996:3}.

We will now show that the local interaction term $D_{kn}$ can be written as a linear transformation of $F_{n+1}$.  First, note that
\begin{align}
\cL_{\ga}(\bfx_k, \bfxp_k, \{ \bfx, \bfxp\}_n,t) &= [v_\gb (\bfx_k,\{\bfx,\bfxp\}_n,t)\pd_{\gb,\bfx_k} + v_\gb (\bfxp_k,\{\bfx,\bfxp\}_n,t)\pd_{\gb,\bfxp_k} ]w_\ga (\bfx_k, \bfxp_k,t) \\
&= \pd_{\gb,\bfx_k} [v_\gb (\bfx_k,\{\bfx,\bfxp\}_n,t) w_\ga (\bfx_k, \bfxp_k,t)] + \pd_{\gb,\bfxp_k} [v_\gb (\bfxp_k,\{\bfx,\bfxp\}_n,t) w_\ga (\bfx_k, \bfxp_k,t)]\\
&= \frac{1}{2n}\sum_{l=1}^n \pd_{\gb,\bfx_k} [ (w_{\gb}(\bfx_k,\bfx_l,t) + w_{\gb}(\bfx_k,\bfxp_l,t))w_\ga (\bfx_k, \bfxp_k,t)]   \\
&\quad +\frac{1}{2n}\sum_{l=1}^n \pd_{\gb,\bfxp_k} [(w_{\gb}(\bfxp_k,\bfx_l,t) + w_{\gb}(\bfxp_k,\bfxp_l,t))w_\ga (\bfx_k, \bfxp_k,t) ],
\end{align}
which gives:
\begin{align}
\cL_{\ga}(\bfx_k - \bfy, \bfxp_k- \bfy, \{ \bfx, \bfxp\}_n,t) &=
 \frac{1}{2n}\sum_{l=1}^n \pd_{\gb,\bfx_k} [ (w_{\gb}(\bfx_k - \bfy,\bfx_l,t) + w_{\gb}(\bfx_k - \bfy,\bfxp_l,t))w_\ga (\bfx_k - \bfy, \bfxp_k - \bfy,t)]   \\
&\quad +\frac{1}{2n}\sum_{l=1}^n \pd_{\gb,\bfxp_k} [(w_{\gb}(\bfxp_k - \bfy,\bfx_l,t) + w_{\gb}(\bfxp_k - \bfy,\bfxp_l,t))w_\ga (\bfx_k - \bfy, \bfxp_k - \bfy,t) ].
\end{align}
It follows from substituting the above to \eqref{eq:dkn} that $D_{kn}$ is given by
\begin{equation}
D_{kn}^{\ga_1\ga_2\cdots\ga_n} (\{\bfx, \bfxp\}_n , t) = \frac{1}{2n}\sum_{l=1}^n \int d\bfy P_{\ga_k \gb}(\bfy) D_{knl}^{\ga_1\ga_2\cdots\ga_{k-1}\gb\cdots\ga_n} (\{\bfx, \bfxp\}_n , \bfy, t),
\end{equation}
with $D_{knl} = D_{knl1} +D_{knl2} + D_{knl3} +D_{knl4}$, and
\begin{align}
D_{knl1}^{\ga_1\cdots\ga_{k-1}\gb\ga_{k+1}\cdots\ga_n} (\{\bfx, \bfxp\}_n , \bfy, t) &= \pd_{\ga_{n+1},\bfx_k} F_{n+1}^{\ga_1\cdots\ga_{k-1}\gb\ga_{k+1}\cdots\ga_{n+1}} (\{\bfX_m\}_{m=1}^{k-1}, \bfx_k - \bfy, \bfxp_k-\bfy, \{\bfX_m\}_{m=k+1}^{n}, \bfx_k - \bfy, \bfx_l), \\
D_{knl2}^{\ga_1\cdots\ga_{k-1}\gb\ga_{k+1}\cdots\ga_n} (\{\bfx, \bfxp\}_n , \bfy, t) &= \pd_{\ga_{n+1},\bfx_k} F_{n+1}^{\ga_1\cdots\ga_{k-1}\gb\ga_{k+1}\cdots\ga_{n+1}} (\{\bfX_m\}_{m=1}^{k-1}, \bfx_k - \bfy, \bfxp_k-\bfy, \{\bfX_m\}_{m=k+1}^{n}, \bfx_k - \bfy, \bfxp_l), \\
D_{knl3}^{\ga_1\cdots\ga_{k-1}\gb\ga_{k+1}\cdots\ga_n} (\{\bfx, \bfxp\}_n , \bfy, t) &= \pd_{\ga_{n+1},\bfxp_k} F_{n+1}^{\ga_1\cdots\ga_{k-1}\gb\ga_{k+1}\cdots\ga_{n+1}} (\{\bfX_m\}_{m=1}^{k-1}, \bfx_k - \bfy, \bfxp_k-\bfy, \{\bfX_m\}_{m=k+1}^{n}, \bfxp_k - \bfy, \bfx_l), \\
D_{knl4}^{\ga_1\cdots\ga_{k-1}\gb\ga_{k+1}\cdots\ga_n} (\{\bfx, \bfxp\}_n , \bfy, t) &= \pd_{\ga_{n+1},\bfxp_k} F_{n+1}^{\ga_1\cdots\ga_{k-1}\gb\ga_{k+1}\cdots\ga_{n+1}} (\{\bfX_m\}_{m=1}^{k-1}, \bfx_k - \bfy, \bfxp_k-\bfy, \{\bfX_m\}_{m=k+1}^{n},\bfxp_k - \bfy, \bfxp_l). 
\end{align}
\end{widetext}

\section{Forcing contribution for gaussian forcing}
\label{sec:gaussianforcing}

We give here a proof of equations \eqref{eq:gforceA} and \eqref{eq:gforceB}, closely following the argument in section II-C-3 of Ref. \onlinecite{article:Procaccia:1996}. We exploit the following mathematical result: if $f_{\ga}(\bfx_1,t_1)$ is a Gaussian stochastic field, the ensemble averages of the form $\avg{f_{\ga}(\bfx_1,t_1)R[f]}$ can be evaluated for any analytic functional  $R[f]$  by the following integral
\begin{widetext}
\begin{equation}
\avg{f_{\ga}(\bfx_1,t_1)R[f]} = \int d\bfx_2 dt_2 \avg{f_{\ga}(\bfx_1, t_1)f_{\gb}(\bfx_2, t_2)} \avg{\frac{\gd R[f]}{\gd f_{\gb}(\bfx_2, t_2)}}.
\label{eq:Novikov}
\end{equation}
We begin the proof by defining the following response functions
\begin{align}
\cG_{\ga\gb} (\bfX, t_1; \bfy, t_2) &= \avg{\frac{\gd w_\ga (\bfX, t_1)}{\gd f_\gb (\bfy, t_2)}}, \\
\cG_{mn}^{\ga_1\cdots\ga_m\gb_1\cdots\gb_n} (\{\bfX\}_m, t, \{\bfy,\gt\}_n) &= \avg{\left[ \prod_{k=1}^n \frac{\gd}{\gd f_{\gb_k} (\bfy_k, \gt_k)}\right] \left[ \prod_{l=1}^m w_{\ga_l} (\bfX_l, t) \right]}.
\end{align}
For the case $t_1=t_2=t$, the response function $\cG_{\ga\gb} (\bfX, t; \bfy, t)$ is given by 
\begin{equation}
\cG_{\ga\gb} (\bfX, t; \bfy, t) = (1/2)[P_{\ga\gb} (\bfx-\bfy) - P_{\ga\gb} (\bfxp-\bfy)].
\end{equation}
This is proved in appendix \ref{sec:onetimeresponse}. Likewise, for the case $m=1$ and $\gt_1 = t$, the response function $\cG_{1n}^{\ga_1\cdots\ga_n\gb}(\{\bfX\}_n,t,\bfy,t)$ is given by
\begin{align}
\cG_{1n}^{\ga_1\cdots\ga_n\gb}(\{\bfX\}_n,t,\bfy,t) &= \avg{\frac{\gd}{\gd f_\gb (\bfy,t)}\left[ \prod_{l=1}^n w_{\ga_l} (\bfX_l,t) \right]} = \sum_{k=1}^n \avg{\left[ \prod_{l=1, l\neq k}^n w_{\ga_l} (\bfX_l,t) \right]\frac{\gd w_{\ga_k}(\bfX_k,t)}{\gd f_\gb (\bfy,t)}}\\
&= \sum_{k=1}^n F_{n-1}^{\ga_1\cdots\ga_{k-1}\ga_{k+1}\cdots\ga_n} (\{\bfX\}_n^k) G_{\ga_k\gb}(\bfX_k, t;\bfy,t).
\end{align}
Here we exploit the fact, first pointed out in Ref. \onlinecite{article:Procaccia:1996}, that the variational derivative $(\gd w_{\ga_k}(\bfX_k,t))/(\gd f_\gb (\bfy,t))$ is not correlated with the velocity differences $w_{\ga_l} (\bfX_l,t)$ because no time is being allowed for interaction to develop a correlation. Using \eqref{eq:Novikov} the correlation between $w_{\ga}(\bfX)$ and $f_{\gb}(\bfy)$ is given by
\begin{align}
\avg{w_{\ga}(\bfX) f_{\gb}(\bfy)} &= \int d\bfz \; \int d\gt\; \avg{\frac{\gd w_{\ga}(\bfX)}{f_{\gc}(\bfz,\gt)}}\avg{f_{\gb}(\bfy,t)f_{\gc}(\bfz,\gt)}= 2\gee \int d\bfz \cG_{\ga\gc}(\bfX,t;\bfz,t) C_{\gb\gc} (\bfy,\bfz),
\end{align}
and it follows that
\begin{align}
Q_{\ga\gb} (\bfX, \bfY) &= 2\gee \int d\bfz \cG_{\ga\gc}(\bfX,t;\bfz,t) [C_{\gb\gc} (\bfy,\bfz) - C_{\gb\gc} (\bfyp,\bfz)]\\
 &=2\gee \int d\bfz [P_{\ga\gc}(\bfx-\bfz) - P_{\ga\gc}(\bfxp-\bfz)][C_{\gb\gc} (\bfy,\bfz) - C_{\gb\gc} (\bfyp,\bfz)].
\end{align}
Using a similar argument for the more general case, we get
\begin{align}
\avg{\left[ \prod_{l=1}^{n-1}w_{\ga_l} (\bfX_l,t) \right] f_{\gb}(\bfy,t)} &=  \int d\bfz \; \int d\gt\; \avg{\frac{\gd}{\gd f_\gc (\bfz,\gt)}\left[ \prod_{l=1}^{n-1} w_{\ga_l} (\bfX_l,t) \right]} \avg{f_{\gb}(\bfy,t)f_{\gc}(\bfz,\gt)}\\
&=  2\gee \int d\bfz \; \cG_{1,n-1}^{\ga_1\cdots\ga_{n-1}\gc}(\{\bfX\}_{n-1}, \bfz) C_{\gb\gc}(\bfy,\bfz), \\
\end{align}
and it follows that
\begin{align}
Q_{kn}^{\ga_1\cdots\ga_{n-1}\gb} (\{\bfX\}_{n-1}, \bfY,t) &= \avg{\left[ \prod_{l=1}^{n-1}w_{\ga_l} (\bfX_l,t) \right](f_{\gb}(\bfy,t) - f_{\gb}(\bfyp,t))}\\
&= 2\gee  \int d\bfz \; \cG_{1,n-1}^{\ga_1\cdots\ga_{n-1}\gc}(\{\bfX\}_{n-1}, \bfz) [C_{\gb\gc}(\bfy,\bfz) - C_{\gb\gc}(\bfyp,\bfz)] \\
&= 2\gee \sum_{l=1}^{n-1} F_{n-2}^{\ga_1\cdots\ga_{l-1}\ga_{l+1}\cdots\ga_{n-1}}(\{\bfX\}_{n-1}^l) \int d\bfz\; \cG_{\ga_l\gc}(\bfX_l,t;\bfz,t)[C_{\gb\gc}(\bfy,\bfz) - C_{\gb\gc}(\bfyp,\bfz)] \\
&= \sum_{l=1}^{n-1} F_{n-2}^{\ga_1\cdots\ga_{l-1}\ga_{l+1}\cdots\ga_{n-1}}(\{\bfX\}_{n-1}^l) Q_{\ga_l\gb} (\bfX_l, \bfY).
\end{align}
This concludes the proof.
\end{widetext}

\section{Evaluation of the one-time response function}
\label{sec:onetimeresponse}

We show how to calculate the one-time response function and use it to show that the ensemble average of the rate of energy injection $\gee_{in} (\bfx)$ is given by $\avg{\gee_{in} (\bfx)}=\gee C_{\ga\ga} (\bfx, \bfx)$. This argument was given previously by McComb \cite{book:McComb:1990}.

We begin with the definition of the response function:
\begin{equation}
G_{\ga\gb} (\bfx_1, t_1; \bfx_2, t_2) = \avg{\frac{\gd u_{\ga} (\bfx_1, t_1)}{\gd f_{\gb}(\bfx_2, t_2)}}.
\end{equation}
We first show that at  equal times $t_1 = t_2$, $G_{\ga\gb}$ is given by
\begin{equation}
G_{\ga\gb} (\bfx_1, t_1; \bfx_2, t_1) = \frac{1}{2} P_{\ga\gb}(\bfx_1-\bfx_2).
\end{equation}
To show this, note that from linearity with respect to forcing
\begin{widetext}
\begin{equation}
u_{\ga} (\bfx, t) = u_{\ga} (\bfx, 0) + \int_0^t ds\; \cA_{\ga} [u_{\ga} (s)] (\bfr) +  \int_0^t ds\; \int d\bfy\; P_{\ga\gb} (\bfx-\bfy) f_{\ga} (\bfy, s),
\end{equation}
where $\cA_{\ga} [u_{\ga} (s)] (\bfr) $ represents the effect of the advection and pressure term. For convenience, we use the abbreviation $g_\ga = \cP_{\ga\gb} f_\ga$. It follows that
\begin{align}
\frac{\gd u_{\ga} (\bfx_1, t_1)}{\gd f_{\gb}(\bfx_2, t_2)} &= 
\int_0^{t_1} ds\; \frac{\gd \cA_{\ga} [u_{\ga} (s)] (\bfx_1)}{\gd f_{\gb}(\bfx_2, t_2)}+ \frac{\gd}{\gd f_{\gb}(\bfx_2, t_2)} \int_0^{t_1} ds\;  \int d\bfy\; P_{\ga\gb} (\bfx-\bfy)  f_{\gb} (\bfy, s)\\
&= \int_{t_2}^{t_1} ds\; \frac{\gd \cA_{\ga} [u_{\ga} (s)] (\bfr)}{\gd f_{\gb}(\bfx_2, t_2)} + \frac{\gd}{\gd f_{\gb}(\bfx_2, t_2)} \int d\bfy \; \int_0^{\infty} dt [  H(t_1-t) P_{\ga\gb}(\bfx_1 - \bfy)] f_{\gb} (\bfy, t) \\
&= 
\int_{t_2}^{t_1} ds\; \frac{\gd \cA_{\ga} [u_{\ga} (s)] (\bfr)}{\gd f_{\gb}(\bfx_2, t_2)} +  H(t_1-t_2) P_{\ga\gb}(\bfx_1 - \bfx_2).
\end{align}
\end{widetext}
with $H(t)$ the Heaviside function, defined as the integral of a delta function:
\begin{align}
H(t) &= \int_0^t \gd (\tau) \; d\tau = \casethree{1}{\text{if } t\in (0,+\infty)}{1/2}{\text{if } t=0}{0}{\text{if } t\in (-\infty,0)}.
\end{align}
For $t_1 = t_2$, the integral of the first term vanishes and $H(0) = 1/2$, therefore it follows that
\begin{equation}
G_{\ga\gb} (\bfx_1, t_1; \bfx_2, t_1) = \frac{1}{2} P_{\ga\gb}(\bfx_1-\bfx_2).
\end{equation}
Also note that in fact there is a discontinuity in the response function and
\begin{equation}
\lim_{\gD t \goto 0^{+}} G_{\ga\gb} (\bfx_1, t +\gD t ; \bfx_2, t) = P_{\ga\gb}(\bfx_1-\bfx_2).
\end{equation}
From this result, it immediately follows that:
\begin{align}
\avg{\gee_{in} (\bfx)} &= \int d\bfx_0\; \int dt_0 \; \avg{f_{\ga}(\bfx, t)f_{\gb}(\bfx_0, t_0)} \\
&=  \int d\bfx_0\;  2\gee C_{\ga\gb} (\bfx, \bfx_0) G_{\ga\gb} (\bfx, t; \bfx_0, t) \\
&=\int d\bfx_0\;  \gee C_{\ga\gb} (\bfx, \bfx_0) P_{\ga\gb} (\bfx- \bfx_0) = \gee C_{\ga\ga} (\bfx, \bfx).
\end{align}

\section{Scaling exponent inequalities}
\label{sec:holder}

We will show here that for an downscale and upscale cascade, correspondingly, the scaling exponents satisfy the inequalities
\begin{align}
\gz_{n+k} &\leq \gz_n + \gz_k \text{ (downscale)},\\
\gz_{n+k} &\geq \gz_n + \gz_k  \text{ (upscale)}.
\end{align}
The first of these inequalities is well-known. 
The key result here is the second inequality, corresponding to the case of an upscale cascade, whose direction reverses, thus giving a convex upward (or flat) dependence of $\gz_n$ as a function of $n$. This should be contrasted with the case of a downscale cascade where the dependence of $\gz_n$ on $n$ is convex downward (or flat). The proof is ``folklore'' and it uses the Schwarz  and \Holder inequalities. An earlier version of this argument was given by Frisch \cite{article:Frisch:1991,book:Frisch:1995}, who in turn cites Feller \cite{book:Feller:1968}. 

Let $p, q \in (1,+\infty)$ with $1/p+1/q=1$, and let $\phi, \psi$ be two random variables with $\phi > 0$ and $\psi > 0$. The \Holder inequality for ensemble averages states that $\avg{\phi\psi} \leq \avg{\phi^p}^{1/p} \avg{\psi^q}^{1/q}$. For $p=q=1/2$ it reduces to the Schwarz inequality: $\avg{\phi\psi}^2 \leq \avg{\phi^2}\avg{\psi^2}$.

We begin by defining  $w (R)$ as the absolute value of the longitudinal velocity difference:
\begin{equation}
w(R) = |(\bfu (\bfx+R\bfe,t)-\bfu (\bfx,t))\cdot\bfe|,
\end{equation}
where $\bfx\in\bbR^d$ is given and $\bfe$ is a unit vector. The proof is based on the following two assumptions:
(a) For a downscale cascade, in the limit $\ell_0\goto\infty$, $w (R)$ scales as $\avg{[w(R)]^n} \sim (R/\ell_0)^{\gz_n}$. 
For an upscale cascade, the same scaling law holds for the limit $\ell_0 \goto 0^+$. 
(b) For finite $\ell_0$ there is a range of scales where the above scaling law continues to hold as an intermediate asymptotic

The proof uses two ``helper'' inequalities that are interesting in themselves. The first ``helper'' inequality is deduced by choosing $\phi = [w(R)]^{(n-1)/2}$ and $\psi =  [w(R)]^{(n+1)/2}$ and employing the Schwarz inequality. It follows that
\begin{align}
\avg{[w(R)]^n}^2 &= \avg{\phi\psi}^2 \leq \avg{\phi^2}\avg{\psi^2} \\
&= \avg{[w(R)]^{n-1}}\avg{[w(R)]^{n+1}},
\end{align}
and therefore
\begin{equation}
\frac{\avg{[w(R)]^n}^2}{\avg{[w(R)]^{n-1}}\avg{[w(R)]^{n+1}}} \sim \fracp{R}{\ell_0}^{2\gz_n-\gz_{n-1}-\gz_{n+1}} < 1.
\end{equation}
To satisfy this inequality under the limit $\ell_0\goto\infty$ we require $2\gz_n-\gz_{n-1}-\gz_{n+1} \geq 0$. Thus we get for a downscale cascade:
\begin{equation}
\gz_{n+1} - \gz_n \leq \gz_n - \gz_{n-1} \text{ (downscale)}.\label{eq:crazyineq1}
\end{equation}
Likewise, for an upscale cascade, the inequality must be satisfied in the limit $\ell_0 \goto 0^+$, which requires $2\gz_n-\gz_{n-1}-\gz_{n+1} \leq 0$. Thus, for an upscale cascade we have
\begin{equation}
\gz_{n+1} - \gz_n \geq \gz_n - \gz_{n-1} \text{ (upscale)}.
\end{equation}

The second ``helper'' inequality is deduced by choosing $\phi = [w(R)]^n$ and $\psi =  [w(R)]^0=1$ and employing the \Holder inequality with $p=(n+1)/n$ and $q=n+1$.  It follows that
\begin{align}
\avg{[w(R)]^n} &\leq \avg{\phi^{(n+1)/n}}^{n/(n+1)} \avg{\psi^{n+1}}^{1/(n+1)} \\
&= \avg{[w(R)]^{n+1}}^{n/(n+1)},
\end{align}
which implies that
\begin{equation}
\frac{\avg{[w(R)]^n}}{\avg{[w(R)]^{n+1}}^{n/(n+1)}} \sim \fracp{R}{\ell_0}^{\gz_n - (n/(n+1))\gz_{n+1}} < 1.
\end{equation}
By similar reasoning, we find that
\begin{align}
\gz_{n+1} &\leq \frac{n+1}{n}\gz_n \text{ (downscale)},\label{eq:crazyineq2}\\
\gz_{n+1} &\geq \frac{n+1}{n}\gz_n \text{ (upscale)}.
\end{align}

Now let us consider the case of a downscale cascade. We assume with no loss of generality that $n>k$ (otherwise for the following step, one may exchange $n$ and $k$). Combining the inequalities \eqref{eq:crazyineq1} and \eqref{eq:crazyineq2} gives
\begin{align}
\gz_{n+k}-\gz_{n} &= \sum_{a=n}^{n+k-1} (\gz_{a+1}-\gz_a)\leq k (\gz_{k+1}-\gz_k) \\
&\leq k\left( \frac{k+1}{k}\gz_k - \gz_k \right) =  \gz_k,
\end{align}
Thus we establish that
\begin{equation}
\gz_{n+k} \leq \gz_n + \gz_k \text{ (downscale)}.
\end{equation}
For the case of the upscale cascade, the exact same argument, with every inequality reversed,  gives
\begin{equation}
\gz_{n+k} \geq \gz_n + \gz_k \text{ (upscale)}.
\end{equation}

\bibliography{references,references-submit}
\bibliographystyle{apsrev}

\end{document}